\documentclass[a4paper,11pt]{article}
\pdfoutput=1 

\usepackage{jcappub} 

\usepackage[T1]{fontenc} 

\title{Quasinormal Modes of Asymptotically (A)dS Black Hole in Lovelock Background}

\author[a]{N. Abbasvandi,}
\author[b,c,1]{M. J. Soleimani,\note{Corresponding author.}}
\author[b,c]{and W.A.T. Wan Abdullah}
\author[a]{Shahidan Radiman}

\affiliation[a]{School of Physics, FST, Universiti Kebangsaan Malaysia\\Bangi, Malaysia}
\affiliation[b]{Physics Department, University of Malaya,\\Bangsar, KL, Malaysia}
\affiliation[c]{NCPP, IPPP, University of Malaya,\\Bangsar, KL, Malaysia}

\emailAdd{Niloofar@siswa.ukm.edu.my}
\emailAdd{msoleima@cern.ch}
\emailAdd{wat@um.edu.my}
\emailAdd{shahidan@ukm.edu.my}

\abstract{We study the quasinormal modes of the massless scalar field in asymptotically (A)dS black holes in Lovelock spacetime by using the sixth order of the WKB approximation. We consider the effects of the second and third order of Lovelock coupling constants on quasinormal frequencies spectrum as well as cosmological constant.}

\begin{document}
\maketitle
\flushbottom

\section{Introduction}
\label{sec:intro}
The existence of large extra dimensions \cite{arkani,antoniadis,arkanni2,randall} which has opened an exciting field of research on commonly predicts the existence of the possibility of production of black holes \cite{argyres,bank,emparan} at particle colliders such as the Large Hadronic Collider (LHC). 
It is expected that the Planck scale is of the order of TeV, and thereby quantum gravity may show itself at hadronic colliders due to the creation of higher dimensional black holes at the LHC. Hence, the higher dimensional black holes have been attracted a lot of interest during the past decade within different theories such as supersymmetric string theory and different extra dimensions scenarios (see for a review \cite{kanti}). In four dimension, considering the absence of higher derivative terms larger than the second order in the Lagrangian and assuming the general coordinate covariance can deduce the Einstein gravity while in higher dimensions, the above assumptions lead to Lovelock theory of gravity \cite{lovelock}. The modified Einstein tensor is nonlinear in the Riemann tensor only if the spacetime has more than 4 dimensions. Therefore, the Lovelock theory is the most natural extension of GR in higher dimensional spacetime which represents interesting scenario to study the corrections of the physics of gravity results at short distance due to the presence of higher order curvature terms in the action. 
\\Moreover, the Einstein gravity is not reliable any more at the energy scale of black hole production. It should be noted that there exist static spherically symmetric black holes in Lovelock gravity \cite{wheeler,boulware,charmousis} and one can consider the black holes produced at the LHC, of this type, supposedly. Based on some pioneering work \cite{takahashi}, the small black holes are unstable in any dimensions.
\\On the other hand, it is important to study quasinormal modes (QNMs) of black holes \cite{vishveshwara,chandrasekhar,kokkotas,nollert,konoplya,berti,konoplya1} to understand properties of black holes in Lovelock gravity. The QNMs of black holes, which depend only on black hole parameters but not depend on the initial perturbation, present a discrete set of complex frequencies corresponding to the solutions of the perturbation equations. In general, the imaginary part of the complex frequencies are associated with the decay timescale of the perturbation while the real part represents the actual frequency of the oscillation. Recently, number of authors had focused on the QNMs of black holes in asymptotically flat spacetime \cite{berti2003,berti20031,berti2004}, asymptotically (anti-)de Sitter (henceforth AdS or dS) spacetimes \cite{mann97,mann98,kalyanarama,danielsson,danielsson99,brady,abdalla}, gravitational perturbation \cite{giammatteo}, scalar perturbation \cite{wang,chakrabarti}, electromagnetic perturbation \cite{lopez}, and Dirac perturbation \cite{cho,jing04,jing05} in different backgrounds. Moreover, black holes in higher dimensional spacetime have extended the investigation of QNMs to higher dimensional spacetime \cite{lopez09,chakrabarti09,kao,cardoso03}. To investigate the QNMs of black holes in higher dimensional spacetime, one direction is to consider black holes in Lovelock as a more general gravitational theory rather than Einstein theory of General Relativity. 
\\In case of the second order Lovelock theory, the so-called Einstein-Gauss-Bonnet theory, some analysis were performed \cite{dotti} and also QNMs had been calculated \cite{chen}. It was shown that there exists scalar mode instability in five dimensions, the tensor mode instability in six dimensions, and no instability in other higher dimensions \cite{beroiz,zhidenko}. Therefore, we need to incorporate higher order of the Lovelock terms in the dimensions higher than six, specially, when we consider black holes at the LHC \cite{rychkov,rizzo}.
Hence, in this paper, we study the QNMs of black holes in the third order of Lovelock theory for eight dimensional spacetime. We are interested here in what will happen with quasinormal spectrum of the black hole if the third order correction term is taken into account. Here, we present a through study of the QNMs of black holes in asymptotically curved spacetime in Lovelock background.
\\In order to compute the QNMs, there are several techniques, such as potential fit \cite{ferrari}, Leaver's continued fraction method \cite{leaver}, Wentzel-Kramers-Brillouin (WKB) \cite{schutz,iyer}, etc. In this paper, we study the QNMs of massless scalar field for asymptotically (A)dS black hole using the 6th order of WKB approximation as the results gets better for higher orders of WKB \cite{konoplya}. We discuss in detail how the QNMs of massless scalar field are influenced by the parameters of higher dimensional black hole spacetime.
\\This paper is organized as follows: In section II, we will give brief information to Lovelock black hole space time. Section III deals with calculation of QNMs for asymptotically (A)dS black hole and detailed analysis performed. In the last section a summary of results and conclusions are given.    

\section{Lovelock Black Hole Spacetime}
Lovelock gravity \cite{lovelock} is one of the natural generalization of Einstein theory in higher dimensional spacetime which maintains the properties of Einstein equation. Thus, we consider third order Lovelock black hole with a cosmological constant term, where the corresponding Lagrangian is given by
\begin{equation} \label{1}
L = \sum\limits_{m = 0}^{{{\left( {D - 1} \right)} \mathord{\left/
			{\vphantom {{\left( {D - 1} \right)} 2}} \right.
			\kern-\nulldelimiterspace} 2}} {{c_m}{\ell _m}}
\end{equation}
where ${{c_m}}$ is an arbitrary constant and ${{\ell _m}}$ is the Euler density of $2k$-dimensional manifolds which is defined by
\begin{equation} \label{2}
{\ell _m} = \frac{1}{{{2^m}}}\delta _{{\mu _1}{\nu _1}...{\mu _m}{\nu _m}}^{{\lambda _1}{\xi _1}...{\lambda _m}{\xi _m}}{R_{{\lambda _1}{\xi _1}}}^{{\mu _1}{\nu _1}}...{R_{{\lambda _m}{\xi _m}}}^{{\mu _m}{\nu _m}}.
\end{equation}
In Eq. (2.2), $\delta _{{\mu _1}{\nu _1}...{\mu _m}{\nu _m}}^{{\lambda _1}{\xi _1}...{\lambda _m}{\xi _m}}$ and ${R_{{\lambda _1}{\xi _1}}}^{{\mu _1}{\nu _1}}$ are the generalized totally antisymmetric Kronecker delta and the Riemann tensor in $D$ dimension respectively. Hereafter, we set ${c_0} =  - 2\Lambda$, ${c_1} = 1$, ${c_2} = {\alpha  \mathord{\left/{\vphantom {\alpha  2}} \right.\kern-\nulldelimiterspace} 2}$, ${c_3} = {\beta  \mathord{\left/{\vphantom {\beta 3}}\right.\kern-\nulldelimiterspace} 3}$, and ${c_m} = 0$ (for $4 \le m$). Therefore, the second and third order of the Lagrangian terms are given by 
\begin{equation} \label{3}
{\ell _2} = {R_{\mu \nu \rho \sigma }}{R^{\mu \nu \rho \sigma }} - 4{R_{\mu \nu }}{R^{\mu \nu }} + {R^2}
\end{equation}
and
\begin{equation} \label{4}
\begin{array}{l}
{\ell _3} = 24{R_{\mu \nu }}^{\rho \sigma }{R_\rho }^\nu {R_\sigma }^\nu  - 24{R_{\mu \nu }}^{\rho \sigma }{R_{\rho \eta }}^{\mu \nu }{R_\sigma }^\eta  + 3R{R_{\mu \nu }}^{\rho \sigma }{R_{\rho \sigma }}^{\mu \nu }\\
+ 8{R_{\mu \nu }}^{\rho \sigma }{R_{\rho \eta }}^{\mu \kappa }{R_{\kappa \sigma }}^{\nu \eta } + 2{R_{\mu \nu }}^{\rho \sigma }{R_{\rho \sigma }}^{\eta \kappa }{R_{\eta \kappa }}^{\mu \nu } + 16{R_\mu }^\nu {R_\nu }^\rho {R_\rho }^\mu \\
- 12R{R_\mu }^\nu {R_\nu }^\mu  + {R^3}.
\end{array}
\end{equation} 
Then, one can derive the Lovelock equation with respect to the metric as follows
\begin{equation} \label{5}
0 = {\vartheta _\lambda }^\xi  = \Lambda {\delta _\mu }^\xi  + G_\lambda ^{\left( 1 \right)\xi } + \alpha G_\lambda ^{\left( 2 \right)\xi } + \beta G_\lambda ^{\left( 3 \right)\xi }
\end{equation}
where $G_\lambda ^{\left( 1 \right)\xi }$ is the Einstein tensor and $G_\lambda ^{\left( 2 \right)\xi }$ $\&$ $G_\lambda ^{\left( 3 \right)\xi }$ are the second and the third order of Lovelock tensors respectively.   
\\It was shown \cite{wheeler,cai} that there exist static exact black hole solutions of the equation (2.5). In this case, the line element takes the following form
\begin{equation} \label{6}
d{s^2} =  - N\left( r \right)d{t^2} + {N^{ - 1}}\left( r \right)d{r^2} + {r^2}{{\bar \gamma }_{ij}}d{x^i}d{x^j}
\end{equation}
where ${{\bar \gamma }_{ij}}$ is the metric of $D-2$ dimensional constant curvature space. In order to determine the function $N\left( r \right)$ by Eq. (2.6), it is convenient to define a new variable $\psi \left( r \right)$ as
\begin{equation} \label{7}
N\left( r \right) = h  - {r^2}\psi \left( r \right)
\end{equation}
which the constant curvature $h  = 1,0$ or $-1$. Then, one can obtain the solution from the following equation
\begin{equation} \label{8}
\frac{\beta }{3}\left( {n - 1} \right)\left( {n - 2} \right)\left( {n - 3} \right)\left( {n - 4} \right){\psi ^3} + \frac{\alpha }{2}\left( {n - 1} \right)\left( {n - 2} \right){\psi ^2} + \psi  - \frac{{2\Lambda }}{{n\left( {n + 1} \right)}} = \frac{\mu }{{{r^{n + 1}}}}
\end{equation}
where $\Lambda$ is cosmological constant and $\mu$ is an integration constant related to $ADM$ mass \cite{myers} as $M = \frac{{2\mu {\pi ^{{{\left( {n + 1} \right)} \mathord{\left/ {\vphantom {{\left( {n + 1} \right)} 2}} \right.
						\kern-\nulldelimiterspace} 2}}}}}{{\Gamma \left( {{{\left( {n + 1} \right)} \mathord{\left/ {\vphantom {{\left( {n + 1} \right)} 2}} \right.
					\kern-\nulldelimiterspace} 2}} \right)}}$, where we used a unit $16\pi G = 1$. Here, $n = D - 2$ which $D$ is the number of spactime dimensions. In the following section we study the QNMs of scalar field by concentrating on the asymptotically de-Sitter spherical solution. 

\section{Quasinormal modes of massless Scalar Field}

Recently much attention has been focused on studying the de Sitter (dS) space and asymptotically dS black holes due to their rich structure. Mellor and Moss \cite{mellor} have been done the first calculations of the fundamental gravitational modes for Reissner-Nordstr\"{o}m de Sitter black holes by using numerical techniques and complemented by Brady \cite{brady} through a numerical time evolution of scalar fields. In this case, some pioneering works have been done through numerical time evolution against WKB predictions \cite{abdalla,zhidenko}. 
\\In this manner, we approach higher dimensional asymptotically dS black holes based on the Lovelock gravity within WKB approximation.
In order to calculate the QNMs of scalar field of asymptotically dS Lovelock black holes, we need to consider the test scalar field in the Lovelock background. In this case, the general perturbation equation for the scalar field $\Phi$ in the curve spacetime, the Klein-Gordon-Fock equation, is given by
\begin{equation} \label{9}
\frac{1}{{\sqrt { - g} }}{\partial _\mu }\left( {\sqrt { - g} {g^{\mu \nu }}{\partial _\nu } - {m ^2}} \right)\Phi  = 0.
\end{equation}
Substituting Eq. (2.7) into Eq. (3.1), and using new variables
\begin{equation} \label{10}
\Phi  = \frac{{\phi \left( r \right)}}{r}{e^{ - i\omega t}}{Y_{L}}\left( {\Omega } \right)
\end{equation} 
where ${Y_{L}}\left( {\Omega } \right)$ represents the coordinates in a d-dimensional sphere. In this way, for a d-dimensional sphere, the spectrum of spherical harmonics is given by
\begin{equation} \label{11}
\Delta {Y_L}\left( \Omega  \right) =  - L\left( {L + d - 1} \right){Y_L}\left( \Omega  \right)
\end{equation} 
with $L \ge 0$
In this manner, going over the tortoise coordinate $d{r^*} = \frac{{dr}}{{N\left( r \right)}}$, Eq. (3.1) can be reduced to the radial perturbation equation in form of the Schr\"{o}dinger wavelike as follows
\begin{equation} \label{12}
\left( {\frac{{{d^2}}}{{d{r^*}^2}} + {\omega ^2} - V\left( {{r^*}} \right)} \right)\Phi  = 0
\end{equation}
where the effective potential has the form
\begin{equation} \label{13}
V\left( r \right) = N\left( r \right)\left( {N\left( r \right)\frac{{n\left( {n - 2} \right)}}{{4{r^2}}} + N'\left( r \right)\frac{n}{{2r}} - {m^2} + \frac{{L\left( {L + d - 1} \right)}}{{{r^2}}}} \right).
\end{equation}
\begin{figure*} \label{1}
	\centering 
	\includegraphics[width=0.5\textwidth]{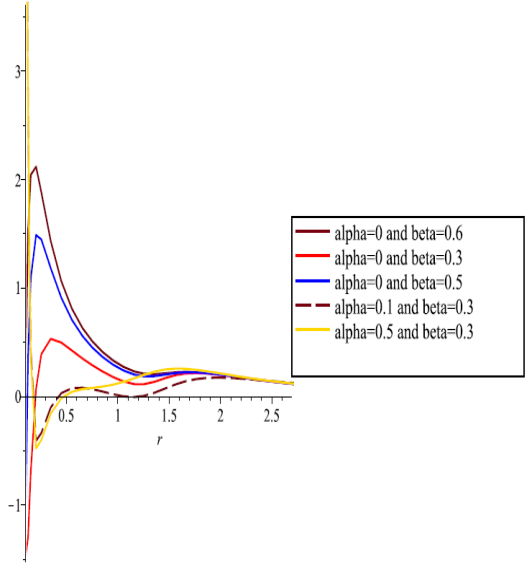}
	\caption{\label{fig:i} The effective potential $V_{n = 6}\left( r \right)$ of Lovelock black holes Vs. $r$ for different $\alpha$, from $\beta = 0$ to $\beta = 0.6$ and $\Lambda=+1$.}
\end{figure*}
\begin{table}[tbp]
	\tiny
	\centering
	\begin{tabular}{|c|c|cc|cc|cc|cc|cc|cc|}
		\hline\hline
		& &\multicolumn{2}{|c|}{$\beta = 0.1$} & \multicolumn{2}{|c|}{$\beta = 0.2$} & \multicolumn{2}{|c|}{$\beta = 0.3$} & \multicolumn{2}{|c|}{$\beta = 0.5$}  & \multicolumn{2}{|c|}{$\beta = 0.6$} \\
		\hline 
		& b & Re $\omega$ & Im $\omega$ & Re $\omega$ & Im $\omega$ & Re $\omega$ & Im $\omega$ & Re $\omega$ & Im $\omega$ & Re $\omega$ & Im $\omega$\\
		\hline 
		& 0 & 0.62747 & -4.22767 & 3.70328 & -3.69759 & 2.90308 & 6.92307 & 116.69226 & 116.72307 & 0.70628 & 1.57482\\
		& 1 & 1.68382 & -6.32683 & 11.51653 & -11.51470 & 4.30552 & 47.85850 & 359.11400 & 359.12402 & 5.07466 & 1.13681\\
		$\alpha = 0.1$ & 2 & 3.61043 & -7.38258 & 23.02915 & -23.02823 & 5.51265 & 104.84458 & 713.16436 & 713.16461 & 11.95982 & 1.26107\\
		& 3 & 6.45219 & -7.85097 & 37.35626 & -37.35570 & 7.69259 & 147.65438 & 1152.69951 & 1152.69967 & 29.93467 & 0.97052\\
		& 4 & 10.17331 & -8.12499 & 54.00666 & -54.00627 & 7.37708 & 254.93175 & 1662.97591 & 1662.97602 & 29.85074 & 1.59724\\
		\hline
		& 0 & 0.65691 & -4.5043 & 2.64944 & -3.57866 & 2.87699 & 6.97348 & 4.63829 & 11.92570 & 0.64971 & -3.94818\\
		& 1 & 1.58164 & -6.95910 & 3.31267 & -28.85864 & 4.29201 & 47.95968 & 7.03509 & 80.83109 & 1.40145 & -6.16714\\
		$\alpha = 0.2$ & 2 & 3.22556 & -8.40246 & 4.19950 & -63.77774 & 5.49690 & 105.04210 & 9.01902 & 176.88528 & 2.92242 & -7.11688\\
		& 3 & 5.54184 & -9.24719 & 4.94501 & -106.40784 & 6.16652 & 184.01753 & 10.65174 & 294.35088 & 5.39035 & -7.24105\\
		& 4 & 8.55680 & -9.75109 & 5.59613 & -155.58225 & 7.35292 & 255.39127 & 12.06887 & 429.9246 & 8.95085 & -7.07675\\
		\hline
		& 0 & 0.45265 & -1.90347 & 2.18762 & 4.22410 & -12.04845 & -0.22195 & 4.63350 & 11.93430 & 1.59781 & 1.75224\\
		& 1 & 0.06281 & 17.61323 & 3.06140 & 30.69120 & -40.59281 & -0.39769 & 7.03197 & 80.85287 & 12.53942 & 12.56003\\
		$\alpha = 0.3$ & 2 & 0.12952 & 38.92878 & 3.90651 & 67.42413 & -82.38232 & -0.52294 & 9.01535 & 176.92825 & 26.93451 & 26.94411\\
		& 3 & 0.16854 & 64.94628 & 4.60799 & 112.31468 &-134.11438 & -0.62252 & 10.64751 & 294.42029 & 44.35362 & 44.35945\\
		& 4 & 0.19817 & 94.95721 & 5.21842 & 164.11392 &-194.09159 & -0.70773 & 12.06413 & 430.02478 & 64.36283 & 64.36685\\
		\hline
		& 0 & 62822.9 & -62822.9 & 0.93415 & 9.04247 & 2.760072 & 7.20144 & 0.65041 & -4.68386 & -2.95489 & -0.39535\\
		& 1 & 219259.5 & -219259.5 & 2.52271 & 35.05546 & 4.21133 & 48.59344 & 18.56050 & -0.5 & -11.72341 &-0.53575\\
		$\alpha = 0.5$ & 2 & $2.694\times {10^5}$ & -269416.03 & 3.25997 & 76.20035 & 5.40097 & 106.30981 & 2.41177 & -9.01734 & -24.47795 & -0.67432\\
		& 3 & $7.313\times {10^5}$ & -731375.4 & 3.85837 & 126.5752 & 6.37936 & 176.89481 & 4.25526 & -9.50568 & -40.19809 & -0.79218\\
		& 4 & $1.059\times {10^6}$ & -1059509.3 & 4.37553 & 184.73798 & 7.22838 & 258.36223 & 7.04655 & -9.27891 & -58.39158 & -0.89558\\
		\hline
		& 0 & -25.91668 & -0.38375 & 1.23685 & 6.41600 & 2.70100 & 7.32459 & 4.61579 & 11.96723 & 0.36144 & -8.66788\\
		& 1 & -85.36000 & -0.66115 & 2.29886 & 36.90289 & 4.17032 & 48.92298 & 7.02032 & 80.93753 & 1.06981 & -8.32339\\
		$\alpha = 0.6$ & 2 & -133.5128 & -1.1191 & 2.98449 & 79.95743 & 5.35206 & 106.9692 & 9.00160 & 177.09592 & 2.00520 & -10.19722\\
		& 3 & -280.19326 & -1.03103 & 3.53677 & 132.69994 & 6.32276 & 177.96482 & 10.63164 & 294.69146 & 3.38128 & -11.16798\\
		& 4 & -405.2205 & -1.17183 & 4.01288 & 193.60781 & 7.16479 & 259.90874 & 12.04632 & 430.41616 & 1.87537 & -32.44592\\
		\hline
	\end{tabular}
	\caption{\label{tab:i} The real and imaginary part of the QNMs of the asymptotically dS Lovelock black holes as a function of $\alpha$ and $\beta$ for $L = 0$ and different overtones for $n = 6$.}
\end{table}
\begin{table}[tbp]
	\tiny
	\centering
	\begin{tabular}{|c|c|cc|cc|cc|cc|cc|cc|}
		\hline\hline
		& &\multicolumn{2}{|c|}{$\alpha = 0.1$} & \multicolumn{2}{|c|}{$\alpha = 0.2$} & \multicolumn{2}{|c|}{$\alpha = 0.3$} & \multicolumn{2}{|c|}{$\alpha = 0.4$}  & \multicolumn{2}{|c|}{$\alpha = 0.5$} \\
		\hline 
		& $\beta$ & Re $\omega$ & Im $\omega$ & Re $\omega$ & Im $\omega$ & Re $\omega$ & Im $\omega$ & Re $\omega$ & Im $\omega$ & Re $\omega$ & Im $\omega$\\
		\hline 
		& 0.15 & 0.16090 & -2.20346 & 0.13271 & -2.42502 & 0.11260 & -2.57164 & 9.66E-02 & -2.61245 & 3.48E-02 & -2.31655\\
		& 0.2 & 0.21916 & -1.76567 & 0.18776 & -1.93927 & 0.16247 & -2.07366 & 0.14101 & -2.15436 & 0.12110 & -2.14495\\
		$\Lambda = 0.1$ & 0.25 & 0.26890 & -1.49164 & 0.25086 & -1.59552 & 0.22363 & -1.70890 & 0.19696 & -1.79726 & 0.17066 & -1.84200\\
		& 0.3 & 0.23072 & -1.49558 & 0.29027 & -1.39784 & 0.28988 & -1.43276 & 0.26647 & -1.50207 & 0.30913 & -1.31932\\
		& 0.35 & 5.71E-02 & -1.92356 & 0.31098 & -1.32357 & 0.28150 & -1.37505 & 0.31790 & -1.30794 & 0.32913 & -1.24082\\
		\hline
		& 0.15 & 0.16016 & -2.20986 & 0.13273 & -2.43207 & 0.11550 & -2.51721 & 0.09951 & -2.54808 & 0.03774 & -2.16958\\
		& 0.2 & 0.20837 & -1.89237 & 0.18816 & -1.94513 & 0.16538 & -2.04438 & 0.14391 & -2.11908 & 0.12400 & -2.10423\\
		$\Lambda = 0.2$ & 0.25 & 0.27022 & -1.49382 & 0.24886 & -1.63245 & 0.22654 & -1.69219 & 0.19987 & -1.77704 & 0.17356 & -1.81797\\
		& 0.3 & 0.23518 & -1.49180 & 0.29135 & -1.39917 & 0.29278 & -1.42257 & 0.26937 & -1.49025 & 0.31203 & -1.31081\\
		& 0.35 & 6.32E-02 & -1.91574 & 0.31998& -1.30556 & 0.28440 & -1.36515 & 0.32081 & -1.29977 & 0.33203 & -1.23351\\
		\hline
		& 0.15 & 0.15828 & -2.22933 &  0.13280 & -2.45282 & 0.11448 & -2.60380 & 1.01E-01 & -2.65819 & 9.60E-02 & -2.45082\\
		& 0.2 & 0.21846 & -1.78569 & 0.16948 & -2.19123 & 0.16561 & -2.10035 & 0.14620 & -2.18708 & 0.13047 & -2.19489\\
		$\Lambda = 0.5$ & 0.25 & 0.27344 & -1.50278 & 0.25517 & -1.61410 & 0.22904 & -1.73088 & 0.20421 & -1.82303 & 0.18155 & -1.87470\\
		& 0.3 & 0.24698 & -1.48427 & 0.30121 & -1.40720 &0.29943 & -1.44968 & 0.27708 & -1.52263 & 0.24950 & -1.58666\\
		& 0.35 & 8.02E-02 & -1.89378 & 0.20934 & -1.57181 &0.30051& -1.37680 & 0.33518 & -1.32084 & 0.32818 & -1.33737\\
		\hline
		& 0.15 & 0.14231 & -2.47853 & 0.13298 & -2.48595 & 0.12448 & -2.59546 & 1.04E-01 & -2.80515 & 9.45E-02 & -2.53967\\
		& 0.2 & 0.21771 & -1.81326 & 0.19056 & -1.99189 & 0.17561 & -2.12311 & 1.56E-01 & -2.20320 & 1.46E-01 &-2.18281\\
		$\Lambda = 1$ & 0.25 & 0.27705 & -1.52253 & 0.25893 & -1.63875 & 0.23904 & -1.76306 & 2.19E-01 & -1.78708 & 2.09E-01 & -1.75297\\
		& 0.3 & 0.26246 & -1.48070 & 0.31104 & -1.42279 & 0.30943 & -1.48362 & 2.89E-01 & -1.48249 & 2.79E-01 & -1.44608\\
		& 0.35 & 1.05E-01 & -1.86224 & 0.22180 & -1.59619 & 0.31051 & -1.41297 & 2.91E-01 & -1.40698 & 2.81E-01 & -1.36802\\
		\hline
		& 0.15 & 0.14915 & -2.47855 & 0.13405 & -2.69464 & 0.13448 & -2.58836 & 1.14E-01 & -2.77849 & 1.04E-01 & -3.00138\\
		& 0.2 & 0.21354 & -2.013071 & 0.19448 & -2.18302 & 0.18561 & -2.14342 & 1.66E-01 & -2.22112 & 1.53E-01 & -2.50442\\
		$\Lambda = 5$ & 0.25 & 0.282077378 & -1.696759795 & 0.267351589 & -1.81459 & 0.24904 & -1.79265 & 2.29E-01 & -1.81820 & 2.14E-01 & -2.11377\\
		& 0.3 & 0.31054 & -1.56275 & 0.33599 & -1.56576 & 0.31943 & -1.51544 & 2.99E-01 & -1.51647 & 3.05E-01 & -1.71646\\
		& 0.35 & 2.16E-01 & -1.75895 & 0.31803 & -1.56098 & 0.32051 & -1.44689 & 3.01E-01 & -1.44335 & 3.24E-01 & -1.54935\\
		\hline
		& 0.15 & 0.14650 & -2.66820 & 0.13440 & -2.87481 & 0.13598 & -2.74365 & 1.21E-01 & -2.83585 & 1.07E-01 & -3.23810\\
		& 0.2 & 0.21018 & -2.18964 & 0.19493 & -2.34943 & 0.18711 & -2.25984 & 1.72E-01 & -2.28249 & 1.58E-01 & -2.68828\\
		$\Lambda = 10$ & 0.25 & 0.27917 & -1.85927 & 0.26706 & -1.97167 & 0.25054 & -1.88170 & 2.36E-01 & -1.87417 & 2.21E-01 & -2.27212\\
		& 0.3 & 0.32339 & -1.68251 & 0.33871 & -1.70743 & 0.32093 & -1.58625 & 3.06E-01 & -1.56597 & 3.01E-01 & -1.93101\\
		& 0.35 & 2.73E-01 & -1.76403 & 0.35136 & -1.63415 & 0.32201 & -1.51778 & 3.07E-01 & -1.49422 & 3.96E-01 & -1.64496\\
		\hline
	\end{tabular}
	\caption{\label{tab:i} The real and imaginary part of the QNMs of the asymptotically dS Lovelock black holes as a function of $\alpha$ and $\beta$ for $L = 2$ and different $\Lambda$ for $n = 6$.}
\end{table}
\begin{table}[tbp]
	\tiny
	\centering
	\begin{tabular}{|c|c|cc|cc|cc|cc|cc|cc|}
		\hline\hline
		& &\multicolumn{2}{|c|}{$\alpha = 0.1$} & \multicolumn{2}{|c|}{$\alpha = 0.2$} & \multicolumn{2}{|c|}{$\alpha = 0.3$} & \multicolumn{2}{|c|}{$\alpha = 0.4$}  & \multicolumn{2}{|c|}{$\alpha = 0.5$} \\
		\hline 
		& $\beta$ & Re $\omega$ & Im $\omega$ & Re $\omega$ & Im $\omega$ & Re $\omega$ & Im $\omega$ & Re $\omega$ & Im $\omega$ & Re $\omega$ & Im $\omega$\\
		\hline 
		& 0.15 & 0.16258 & -2.19093 & 0.13759 & -2.35361 & 0.11748 & -2.48442 & 1.01E-01 & -2.50952 & 3.97E-02 & -2.08988\\
		& 0.2 & 0.21954 & -1.75718 & 0.19264 & -1.90056 & 0.16735 & -2.02694 & 0.14589 & -2.09806 & 0.12597 & -2.08012\\
		$\Lambda = -0.1$ & 0.25 & 0.26378 & -1.52824 & 0.25574 & -1.57292 & 0.22851 & -1.68246 & 0.20184 & -1.76520 & 0.17554 & -1.80390\\
		& 0.3 & 0.22734 & -1.53051 & 0.29515 & -1.38153 & 0.29476 & -1.41683 & 0.27135 & -1.48353 & 0.31401 & -1.30613\\
		& 0.35 & 5.81E-02 & -1.82877 & 0.31586 & -1.30947 & 0.28637 & -1.35965 & 0.32278 & -1.29528 & 0.33401 & -1.22957\\
		\hline
		& 0.15 & 0.16355 & -2.18487 & 0.14566 & -2.15445 & 0.13332 & -2.13074 & 1.17E-01 & -2.10425 & 5.56E-02 & -1.35367\\
		& 0.2 & 0.21975 & -1.75356 & 0.20109 & -1.77535 & 0.18319 & -1.80910 & 0.16173 & -1.84435 & 0.14181 & -1.79279\\
		$\Lambda = -0.2$ & 0.25 & 0.28316 & -1.39387 & 0.26180 & -1.51750 & 0.24435 & -1.54148 & 0.21768 & -1.60092 & 0.19138 & -1.61384\\
		& 0.3 & 0.24812 & -1.37784 & 0.30429 & -1.31018 & 0.31059 & -1.31946 & 0.02587 & -1.52560 & 0.32984 & -1.21976\\
		& 0.35 & 7.61E-02 & -1.47225 & 0.33292 & -1.22787 & 0.30221 & -1.26257 & 0.33862 & -1.21166 & 0.34984 & -1.15160\\
		\hline
		& 0.15 & 0.16708 & -2.16805 &  0.14160 & -2.36663 & 0.12328 & -2.49402 & 1.09E-01 & -2.53016 & 1.05E-01 & -2.33447\\
		& 0.2 & 0.22726 & -1.75782 & 0.17827 & -2.13568 & 0.17441 & -2.04816 & 0.15500 & -2.12343 & 0.13927 & -2.12356\\
		$\Lambda = -0.5$ & 0.25 & 0.28224 & -1.48916 & 0.26397 & -1.59582 & 0.23784 & -1.70627 & 0.21301 & -1.79174 & 0.19035 & -1.83731\\
		& 0.3 & 0.25578 & -1.46987 & 0.31001 & -1.39750 & 0.30823 & -1.43872 & 0.28588 & -1.50856 & 0.25830 & -1.56892\\
		& 0.35 & 8.90E-02 & -1.81192 & 0.21813 & -1.55140 & 0.30930 & -1.36795 & 0.34398 & -1.31431 & 0.33697 & -1.33027\\
		\hline
		& 0.15 & 0.17625 & -2.14951 & 0.14833 & -2.39261 & 0.10548 & -2.40314 & 1.04E-01 & -2.45757 & --- & ---\\
		& 0.2 & 0.22187 & -1.75950 & 0.20592 & -1.96149 & 0.19976 & -2.03511 & 1.95E-01 & -2.05448 & --- & ---\\
		$\Lambda = -1$ & 0.25 & 0.29241 & -1.52577 & 0.27428 & -1.63570 & 0.26319 & -1.72931 & 2.59E-01 & -1.73843 & --- & ---\\
		& 0.3 & 0.27782 & -1.48642 & 0.32640 & -1.43038 & 0.33358 & -1.47723 & 3.29E-01 & -1.48084 & --- & ---\\
		& 0.35 & 1.20E-01 & -1.82671 & 0.24595 & -1.57647 & 0.33466 & -1.41170 & 3.30E-01 & -1.41437 & --- & ---\\
		\hline
		
	\end{tabular}
	\caption{\label{tab:i} The real and imaginary part of the QNMs of the asymptotically AdS Lovelock black holes as a function of $\alpha$ and $\beta$ for $L = 2$ and different $\Lambda$ for $n = 6$.}
\end{table}
\begin{figure*} \label{2}
	\centering 
	\includegraphics[width=0.5\textwidth]{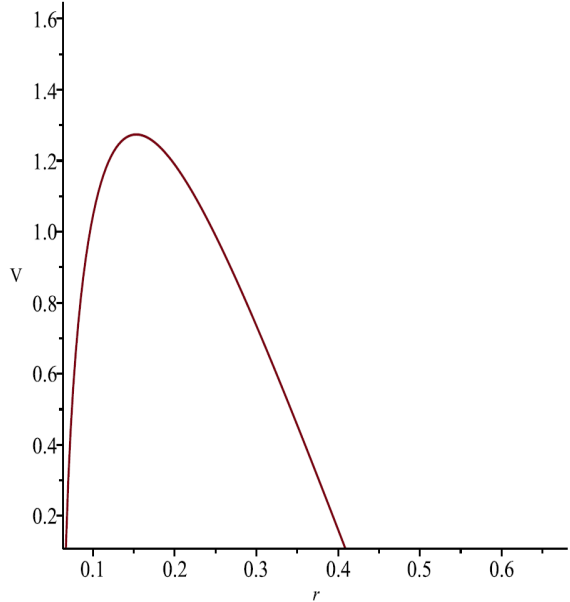}
	\caption{\label{fig:i} The effective potential of scalar perturbation in scalar field for $n=6$, $\alpha=0.7$, $\beta=0.3$, $L=2$, and $\mu=50$ .}
\end{figure*}
\begin{figure*}[tbp] \label{3}
	\centering 
	\includegraphics[width=.35\textwidth,origin=a,trim=-90 70 200 100,]{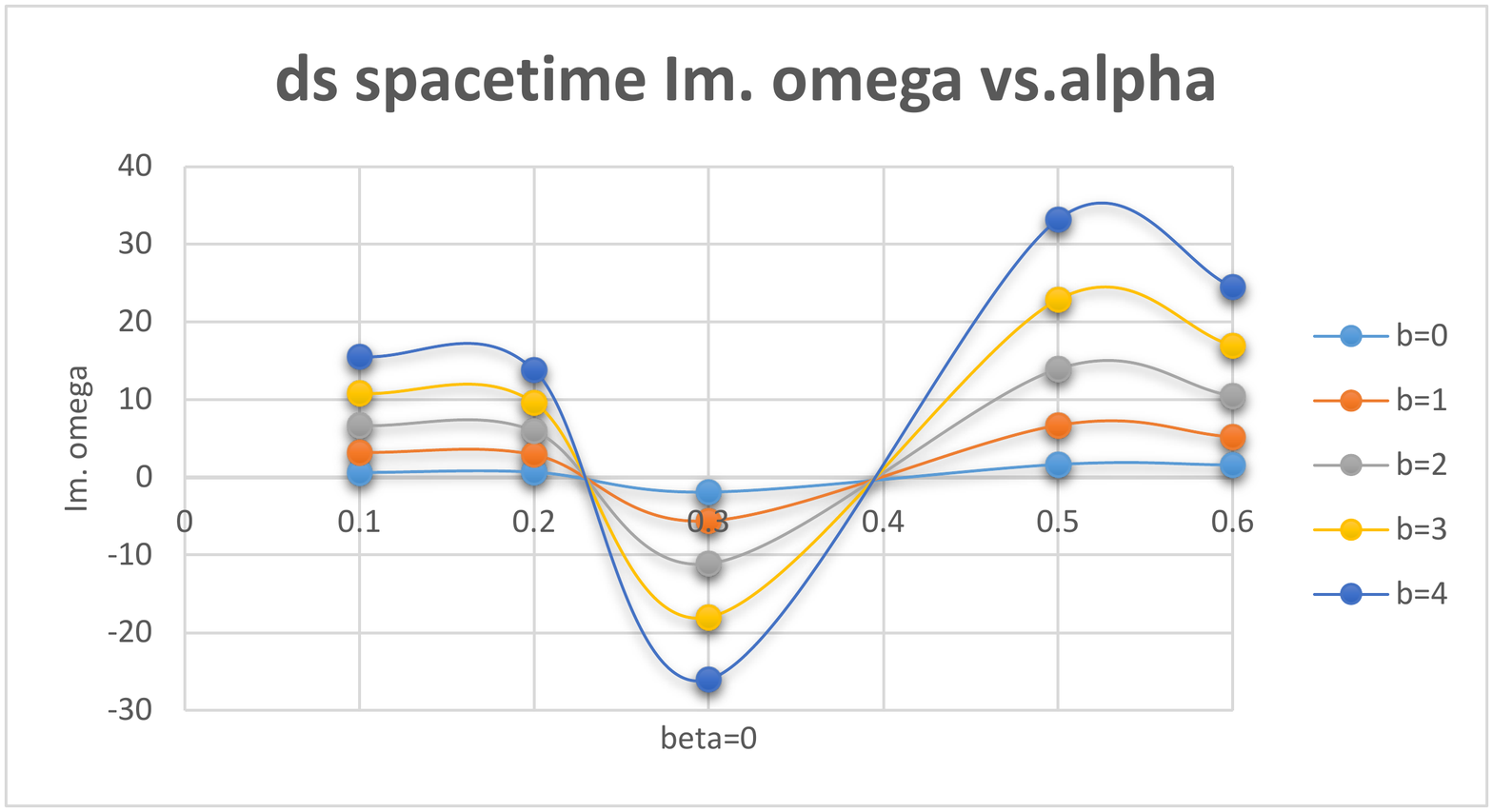}
	\hfill 
	\includegraphics[width=.35\textwidth,origin=a,trim=-90 70 200 100,]{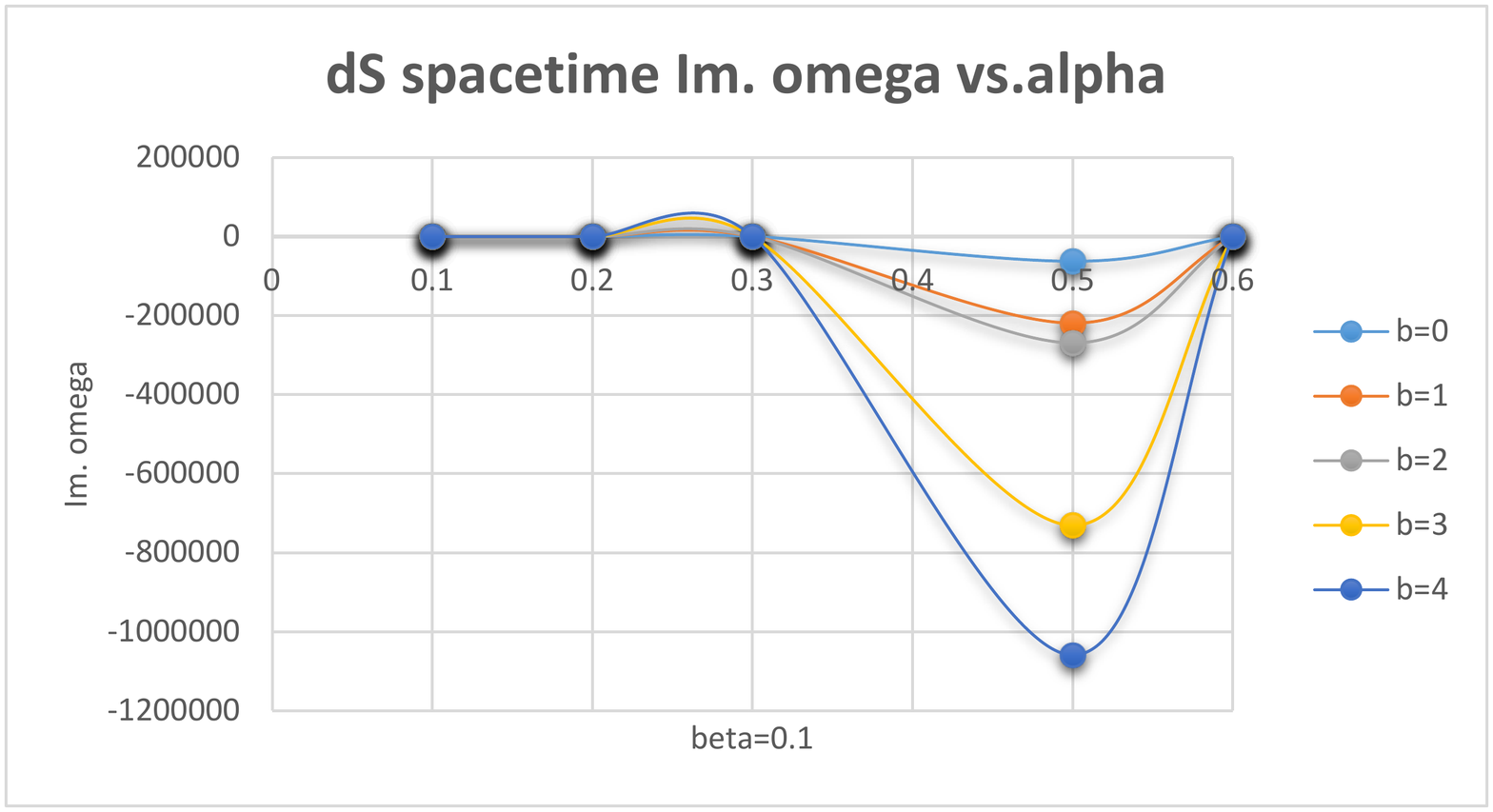}
	\hfill
	\includegraphics[width=.35\textwidth,origin=a,trim=-90 70 200 100, angle=0]{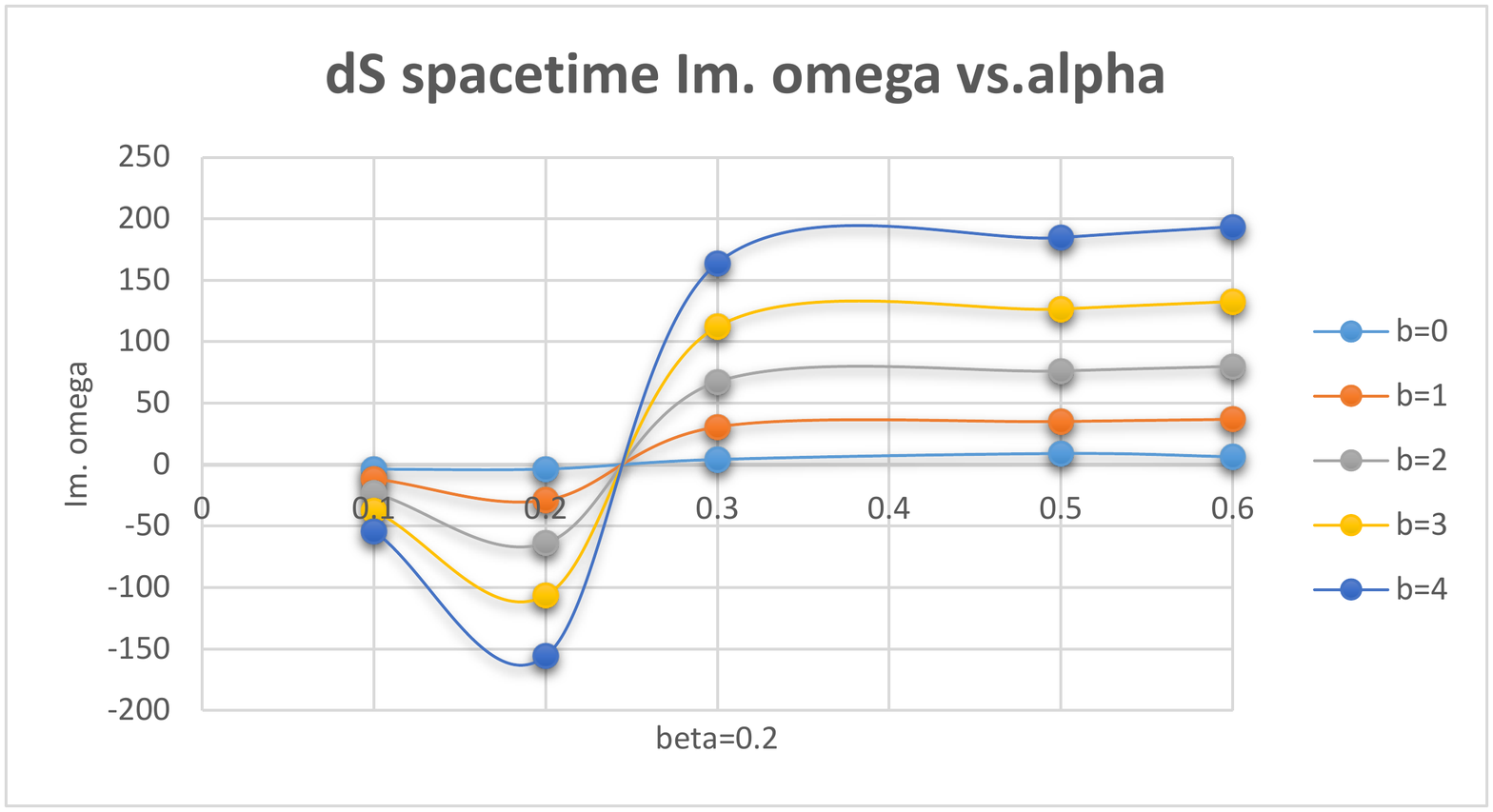}
	\hfill
	\includegraphics[width=.35\textwidth,origin=a,trim=-90 70 200 60, angle=0]{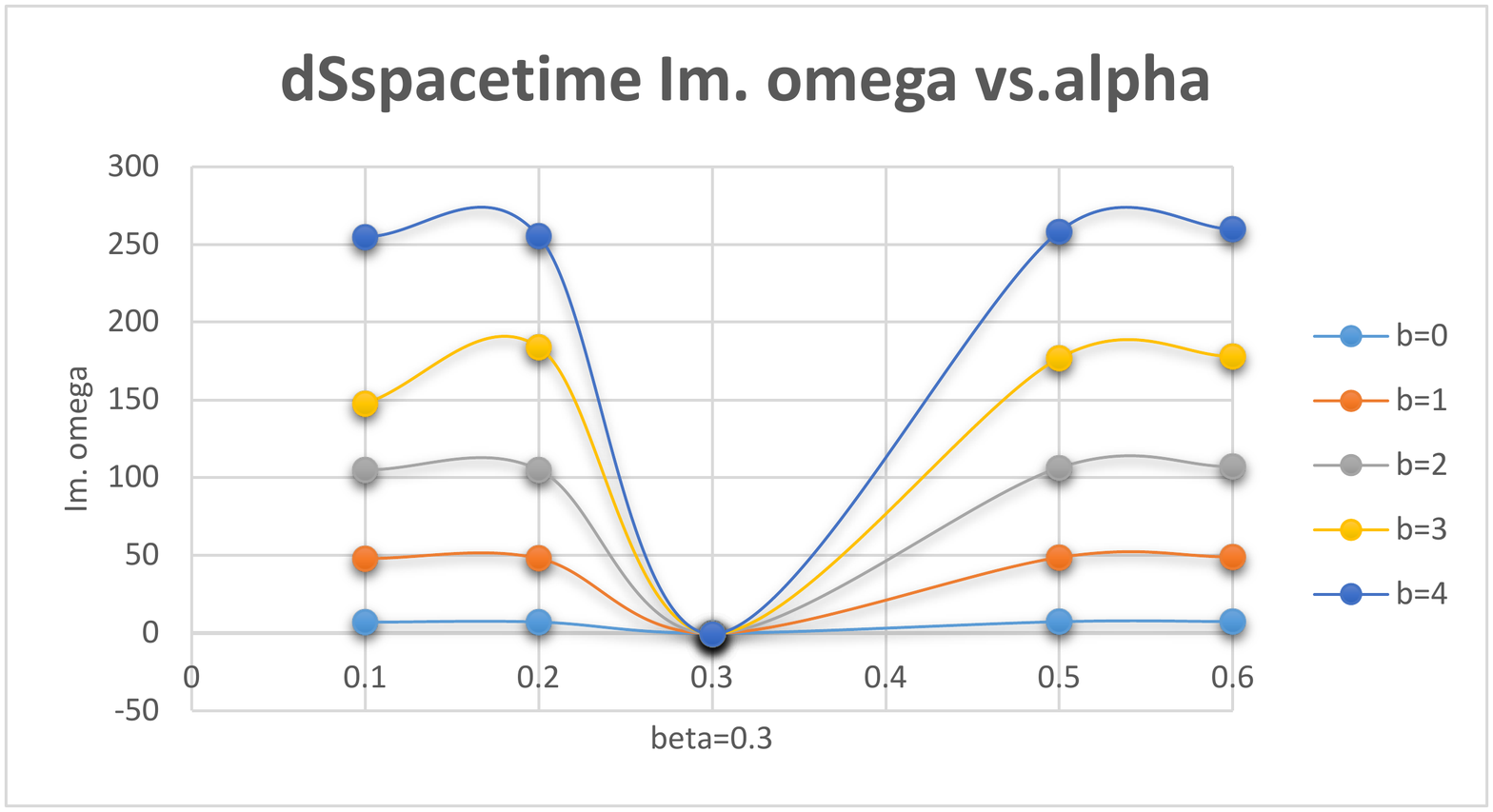}
	\hfill
	\includegraphics[width=.35\textwidth,origin=a,trim=-90 70 200 60, angle=0]{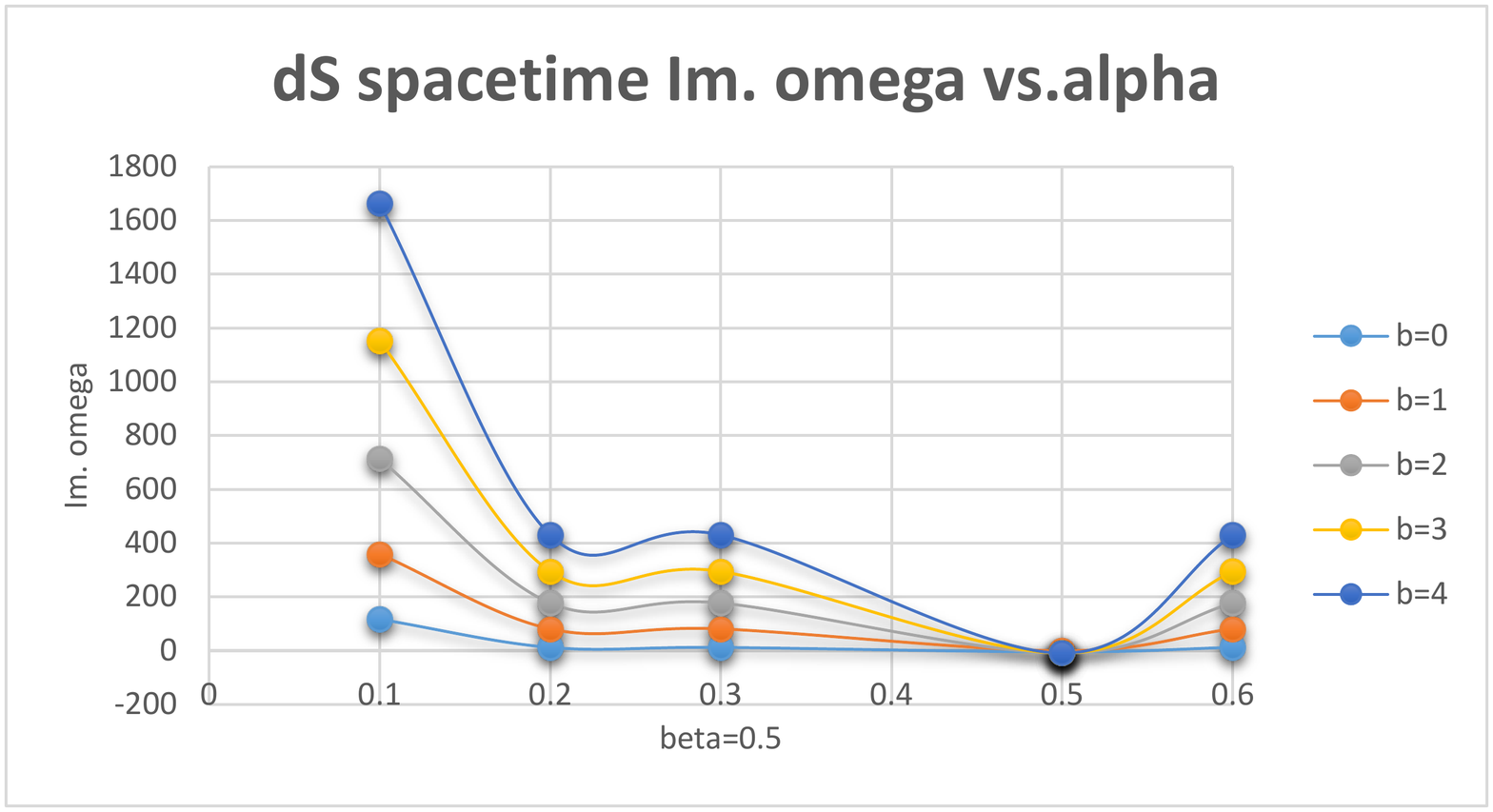}
	\hfill
	\includegraphics[width=.35\textwidth,origin=a,trim=-90 70 200 60, angle=0]{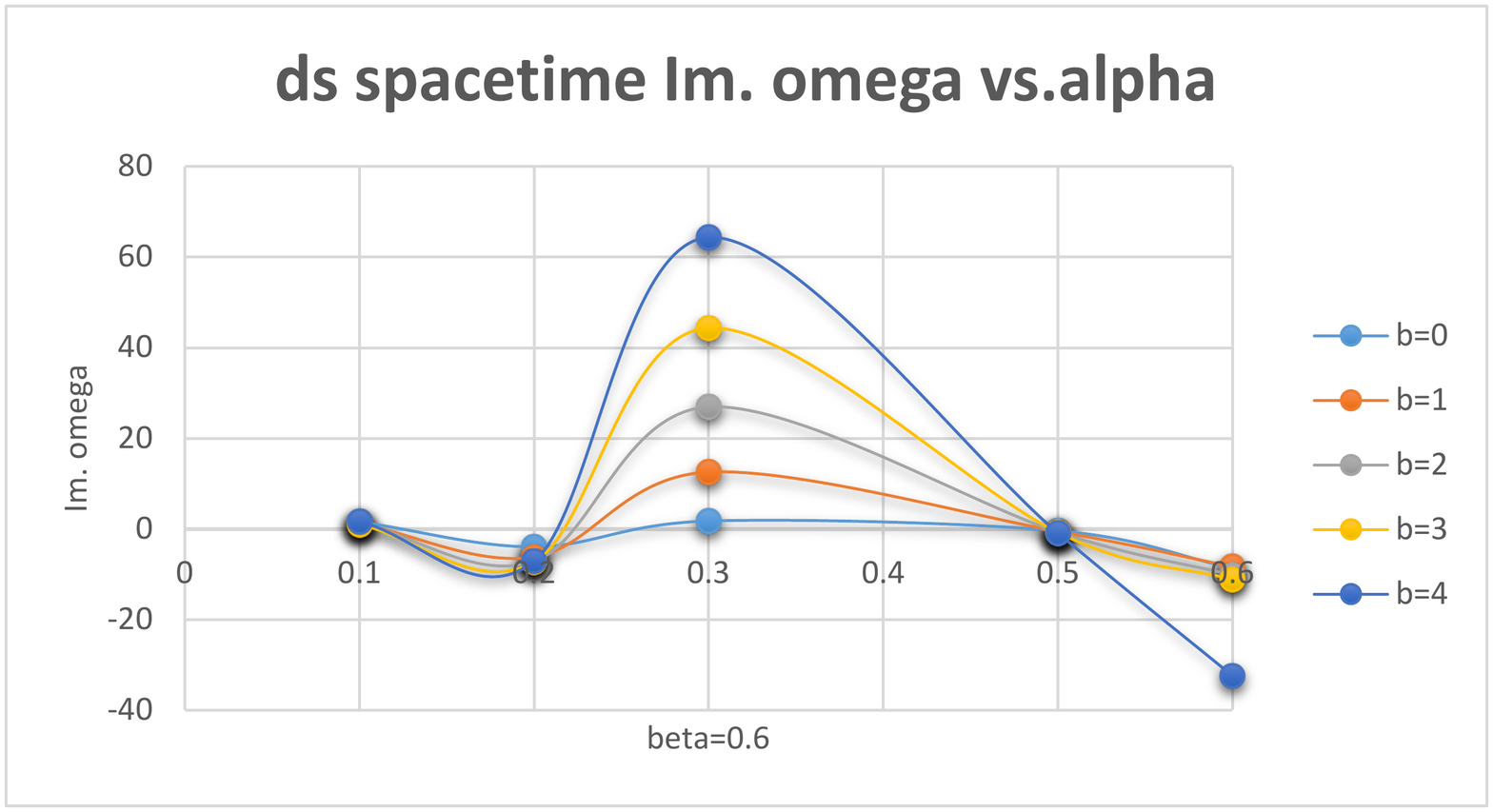}
	\hfill
	\includegraphics[width=.35\textwidth,origin=a,trim=-90 70 200 60, angle=0]{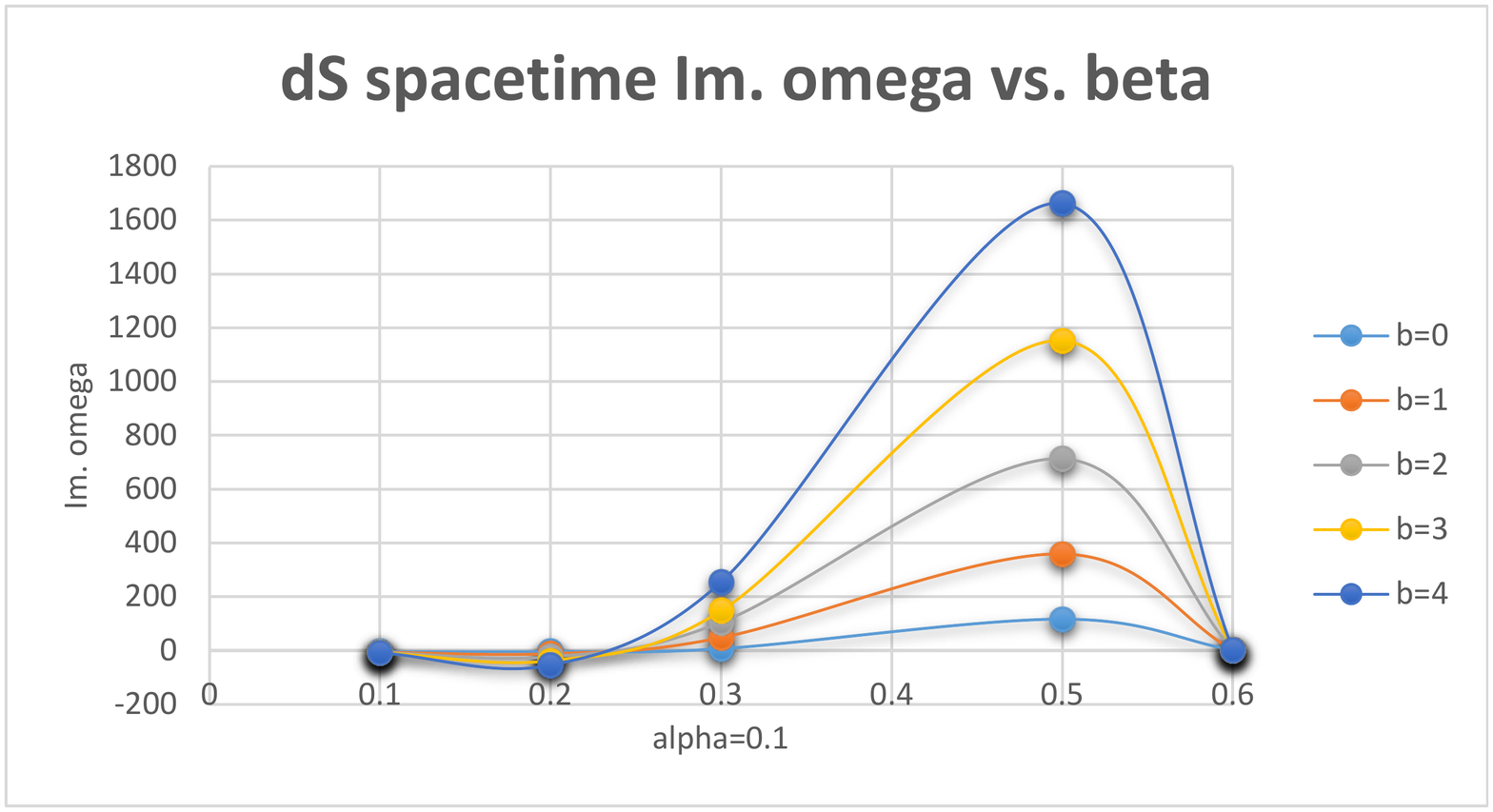}
	\hfill
	\includegraphics[width=.35\textwidth,origin=a,trim=-90 70 220 60, angle=0]{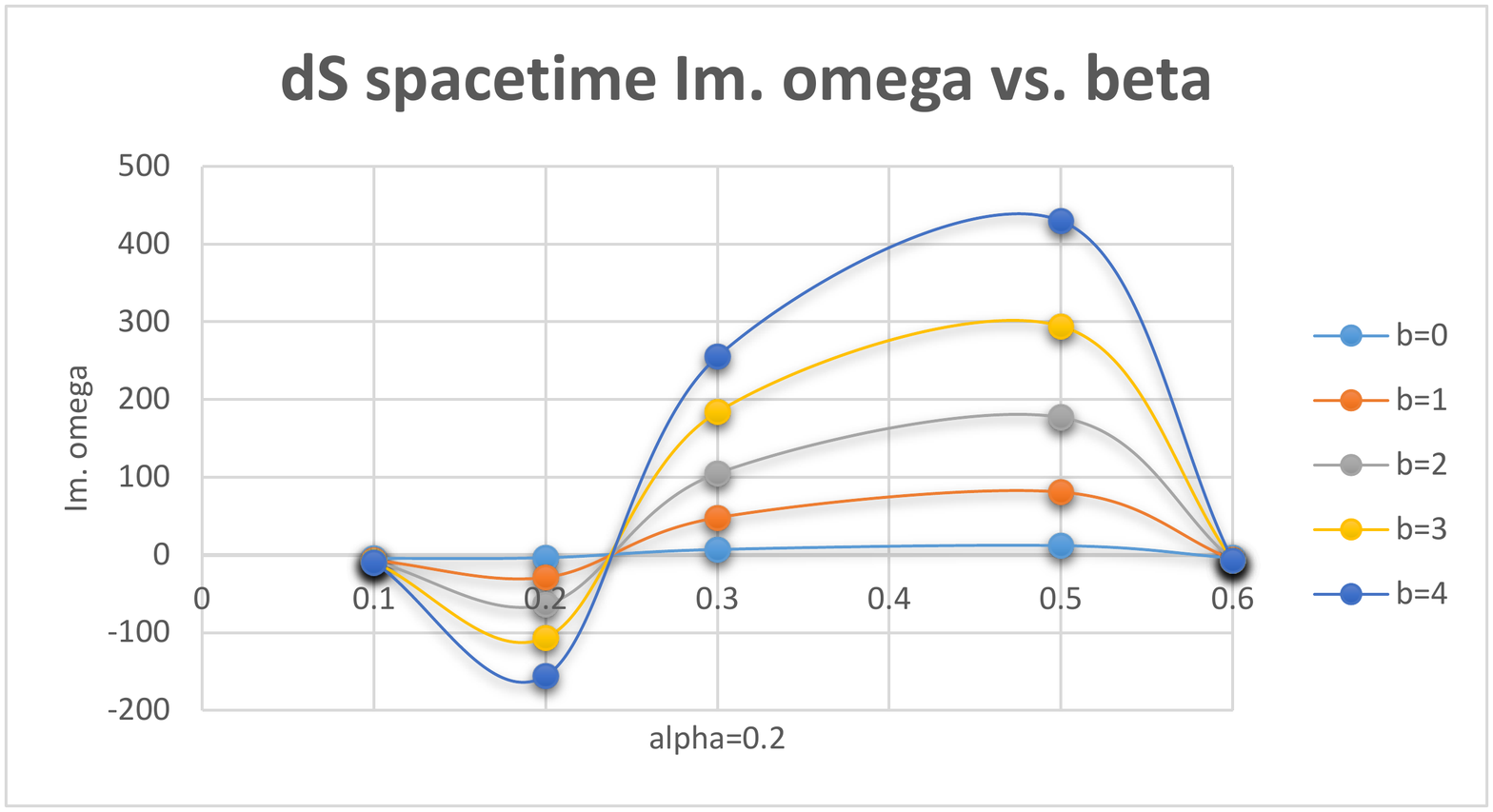}
	\hfill
	\includegraphics[width=.35\textwidth,origin=a,trim=-90 70 220 60, angle=0]{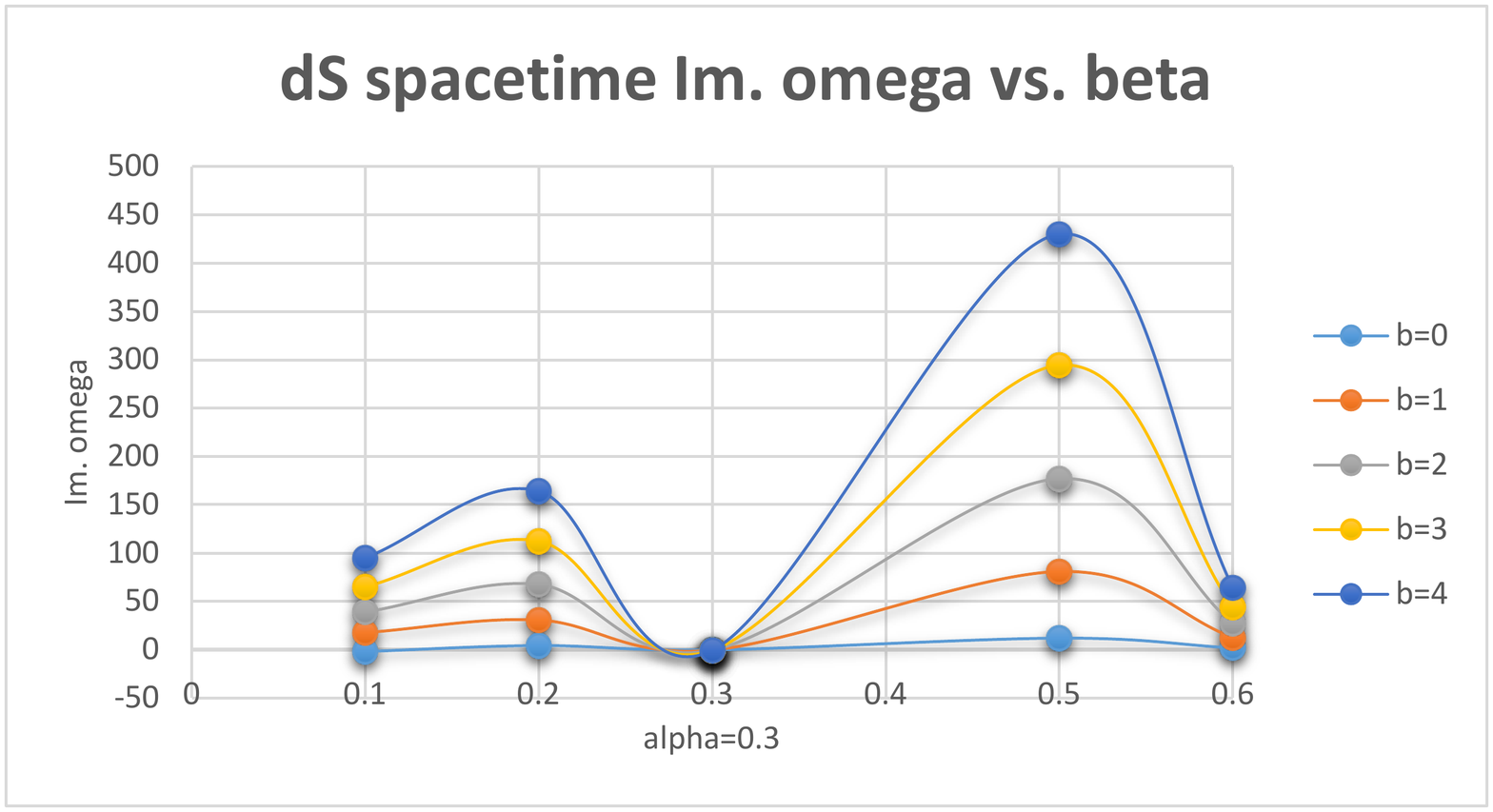}
	\hfill
	\includegraphics[width=.35\textwidth,origin=a,trim=-90 70 200 60, angle=0]{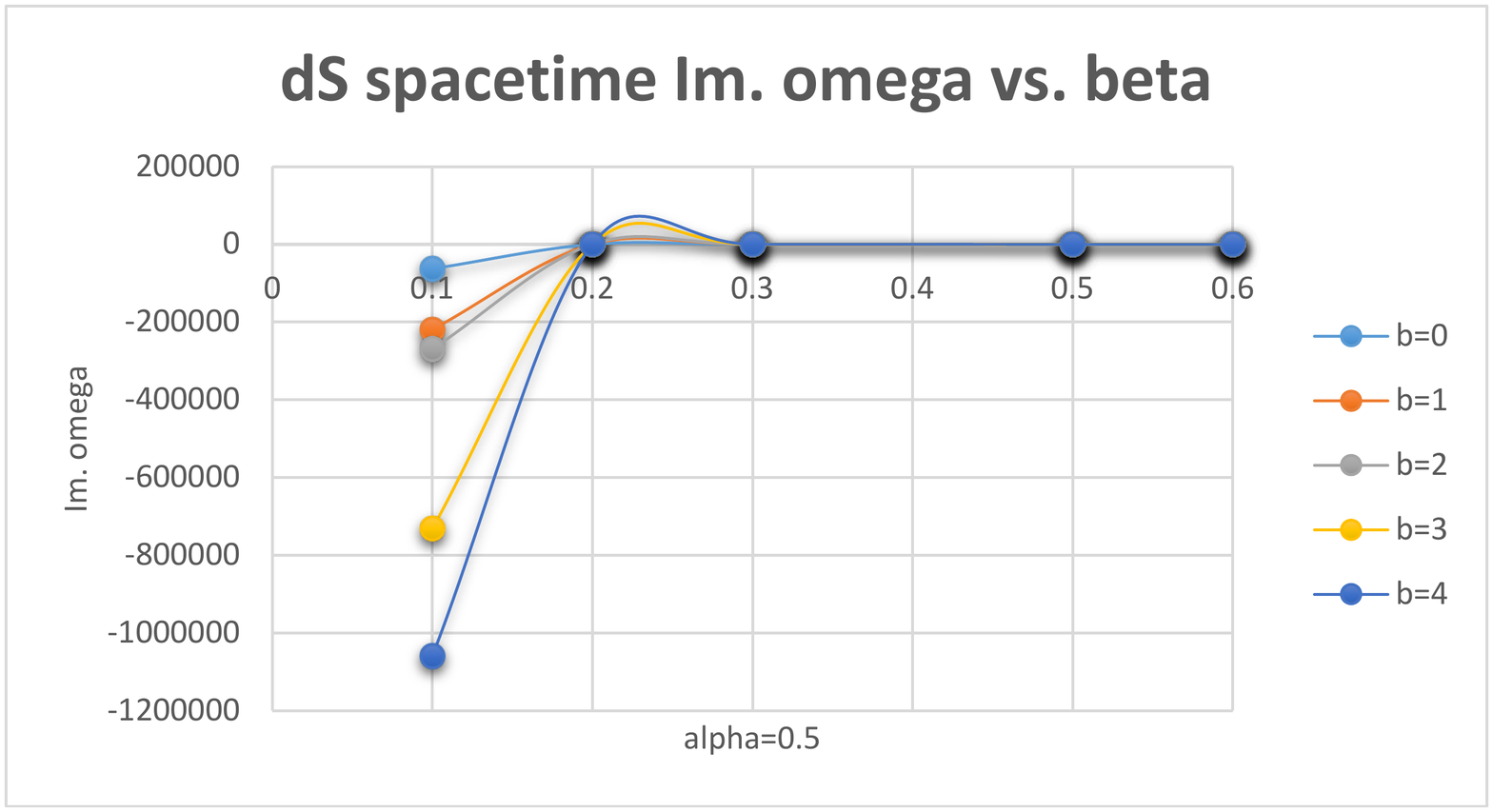}
	\hfill
	\includegraphics[width=.35\textwidth,origin=a,trim=-90 70 200 60, angle=0]{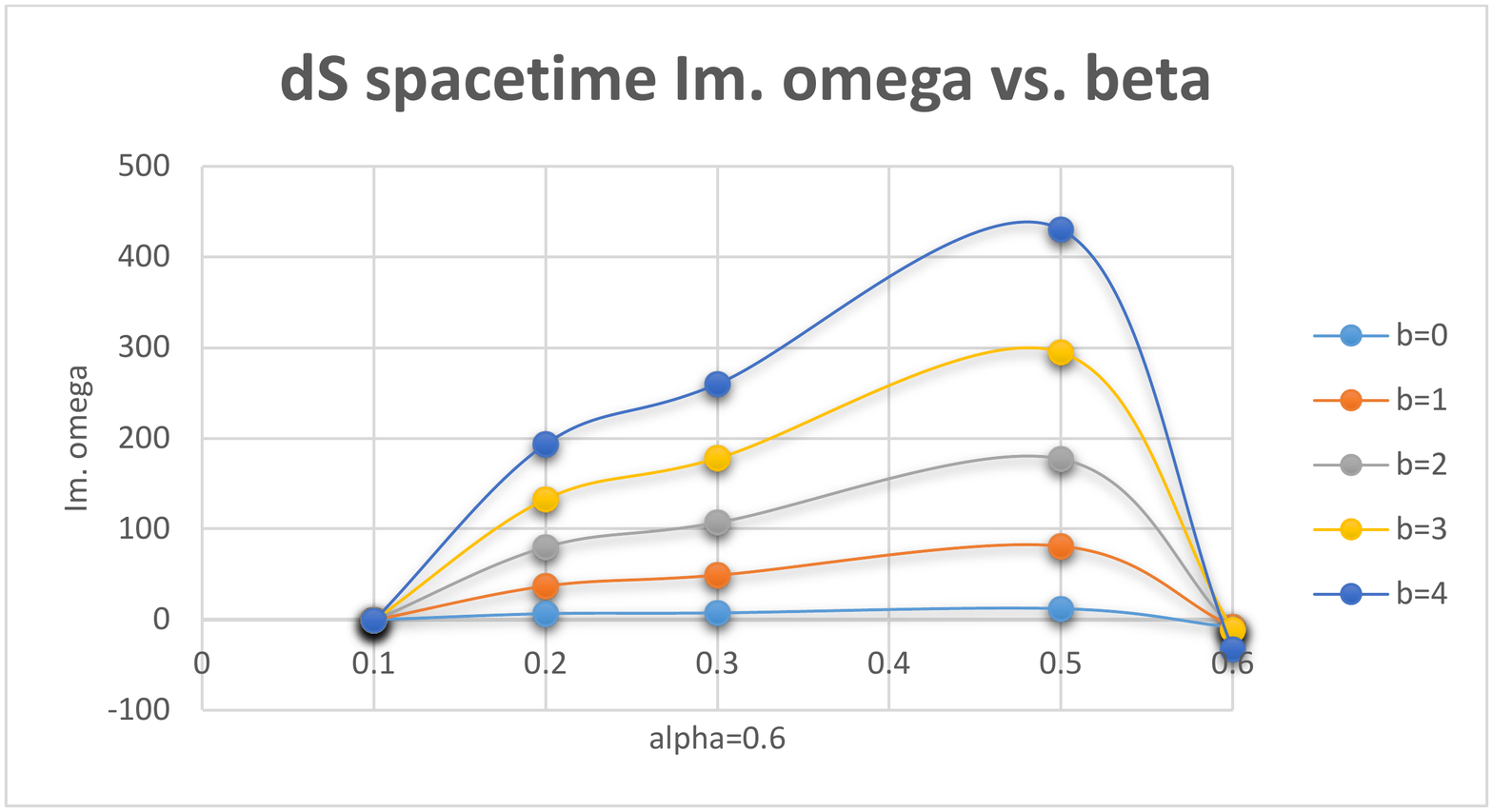}
	\hfill
	
	\caption{\label{fig:i} The imaginary part of the QNM of asymptotically dS black holes as function of $\alpha$ and $\beta$ for different overtone from $b = 0$ to $b = 4$.}
\end{figure*}
\begin{figure}[tbp] \label{4}
	\centering 
	\includegraphics[width=.35\textwidth,origin=a,trim=-90 70 200 100,]{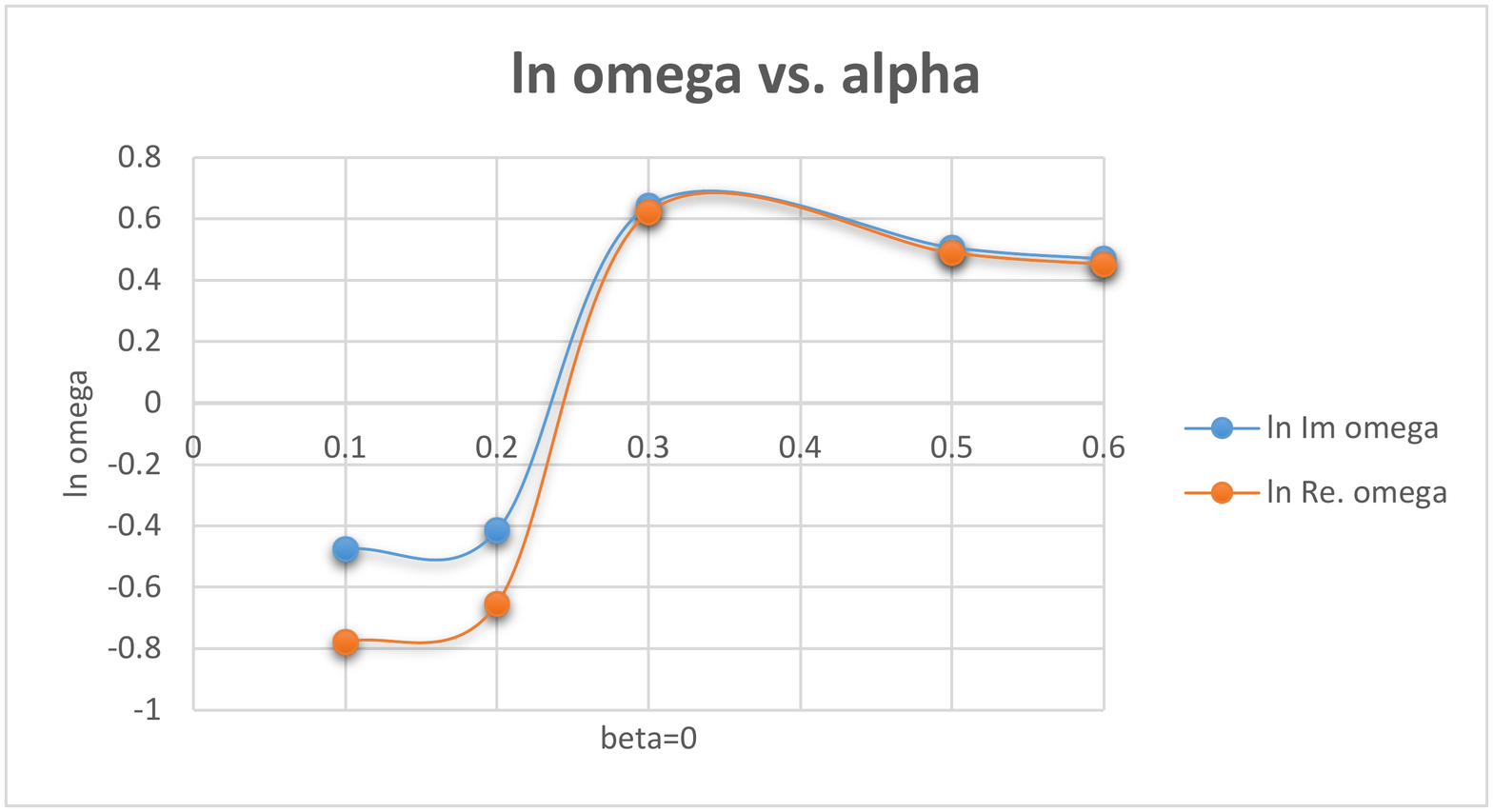}
	\hfill 
	\includegraphics[width=.35\textwidth,origin=a,trim=-90 70 200 100,]{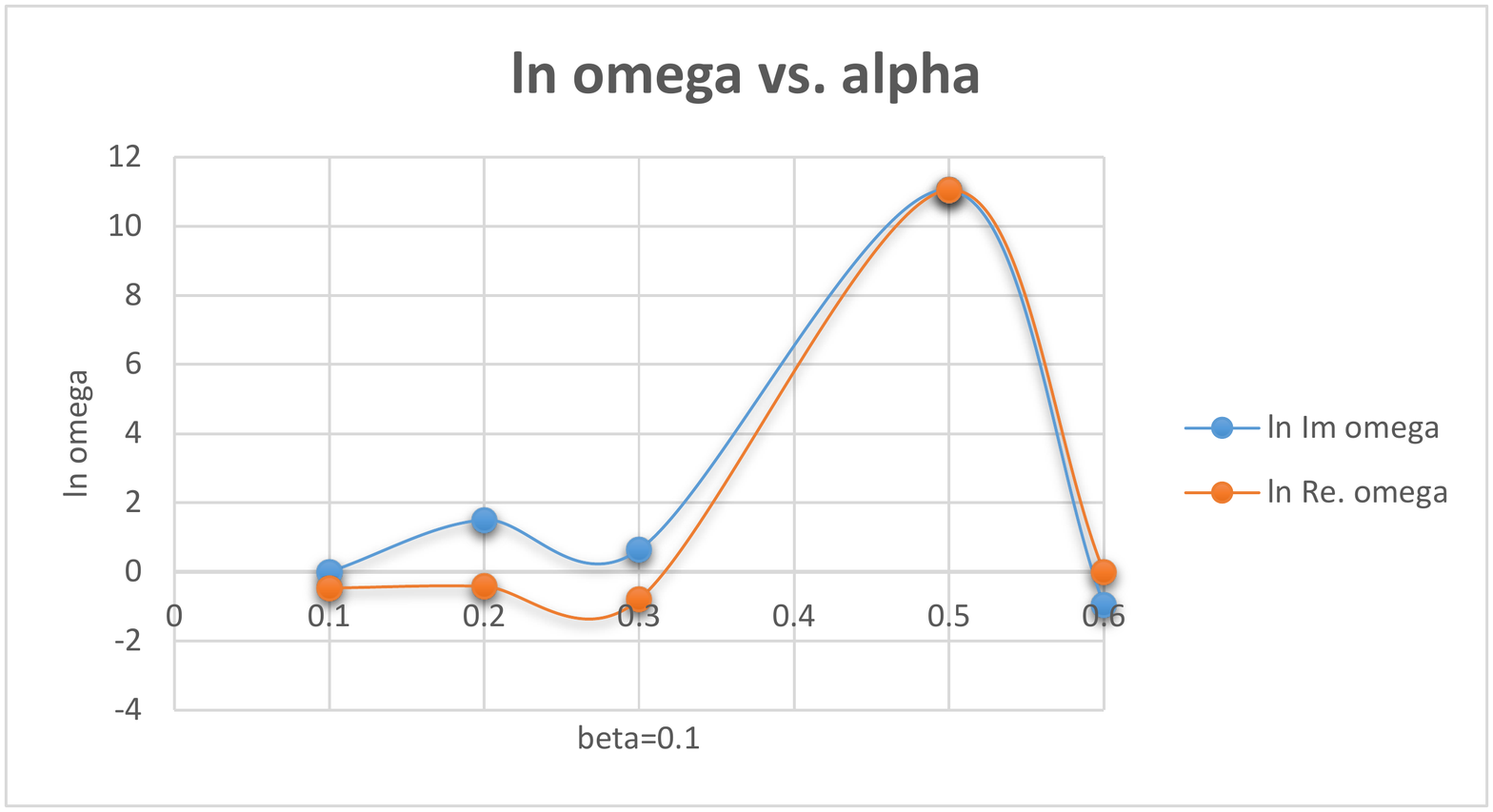}
	\hfill
	\includegraphics[width=.35\textwidth,origin=a,trim=-90 70 200 100, angle=0]{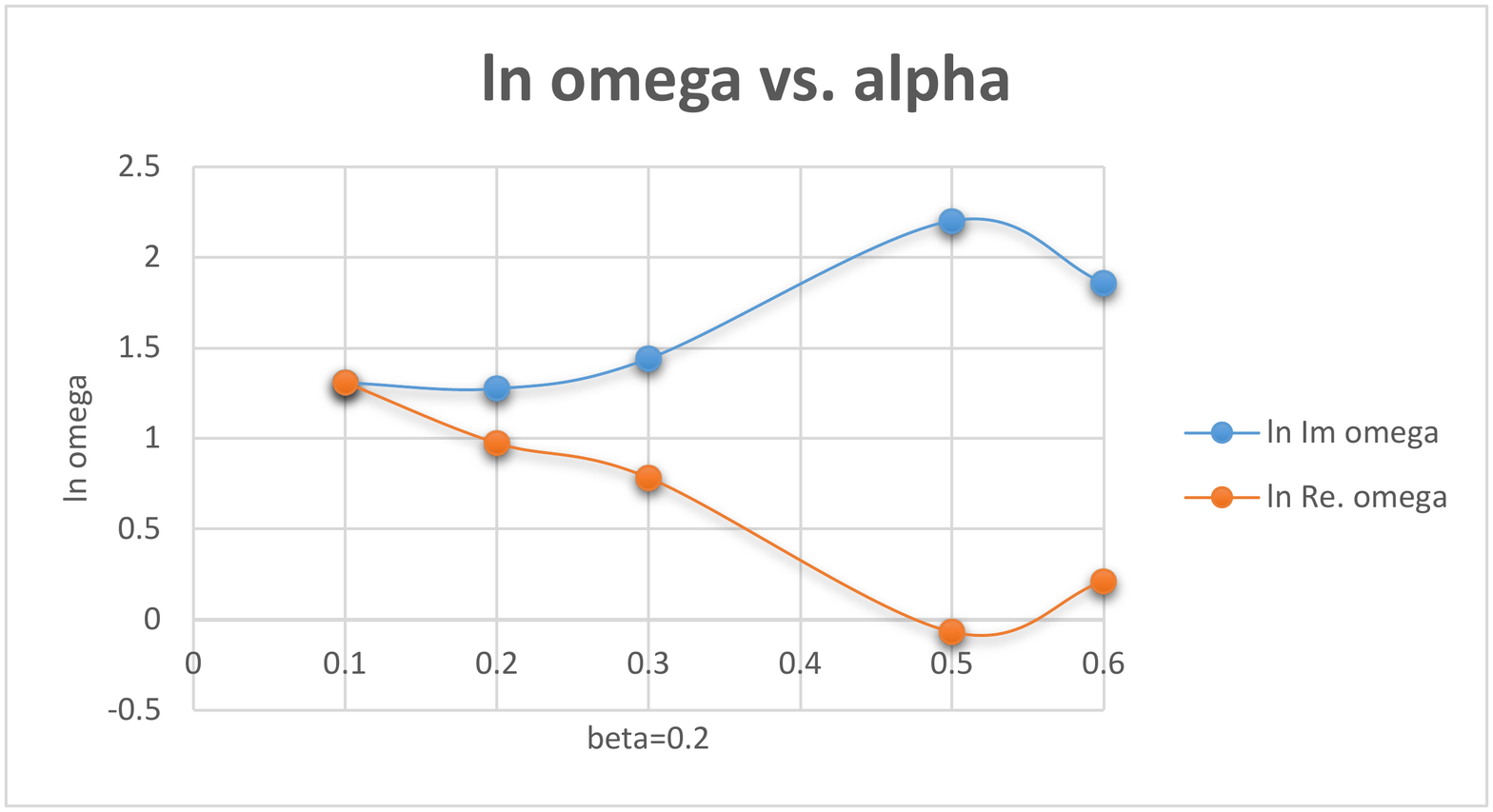}
	\hfill
	\includegraphics[width=.35\textwidth,origin=a,trim=-90 70 200 60, angle=0]{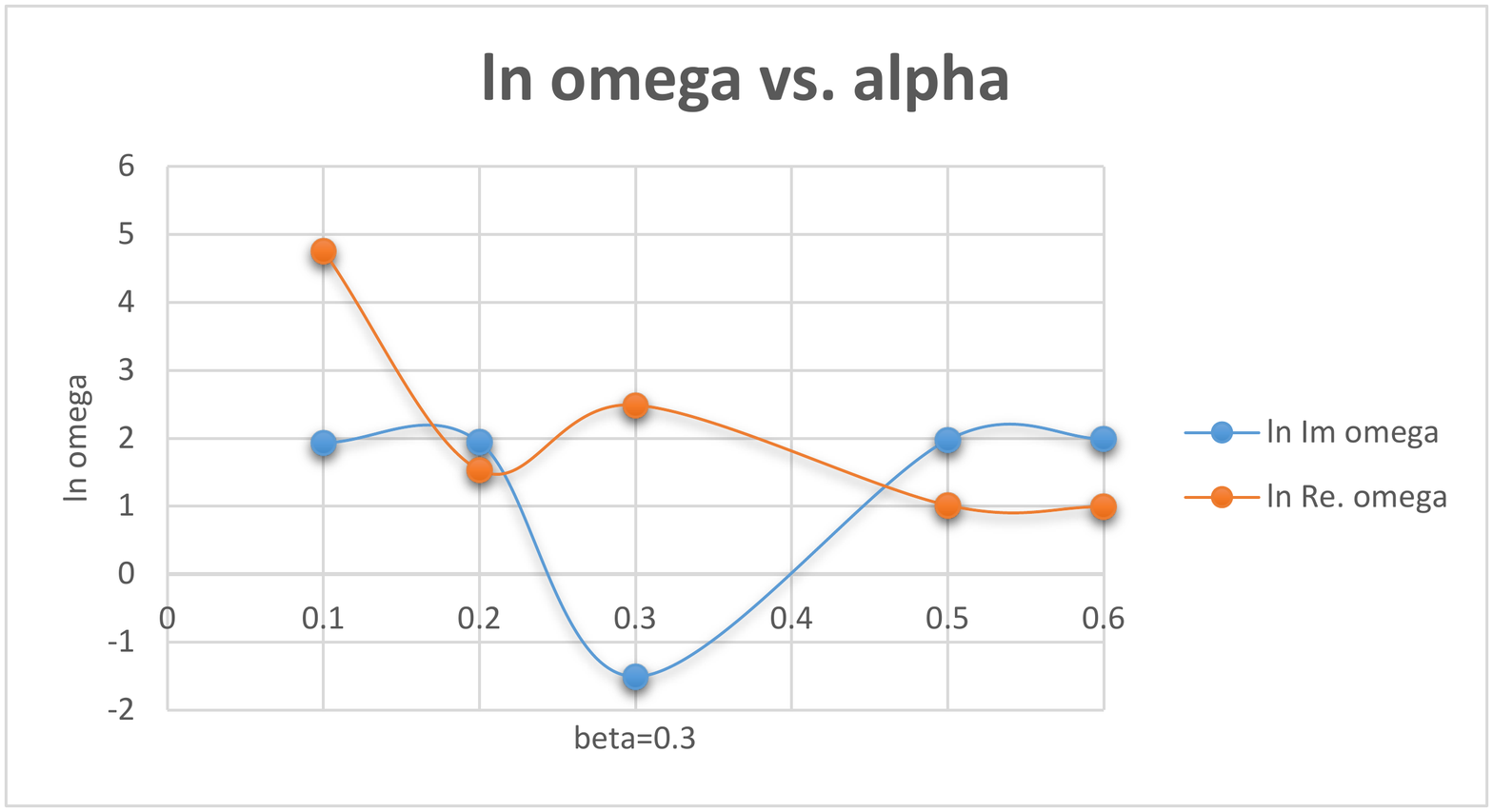}
	\hfill
	\includegraphics[width=.35\textwidth,origin=a,trim=-90 70 200 60, angle=0]{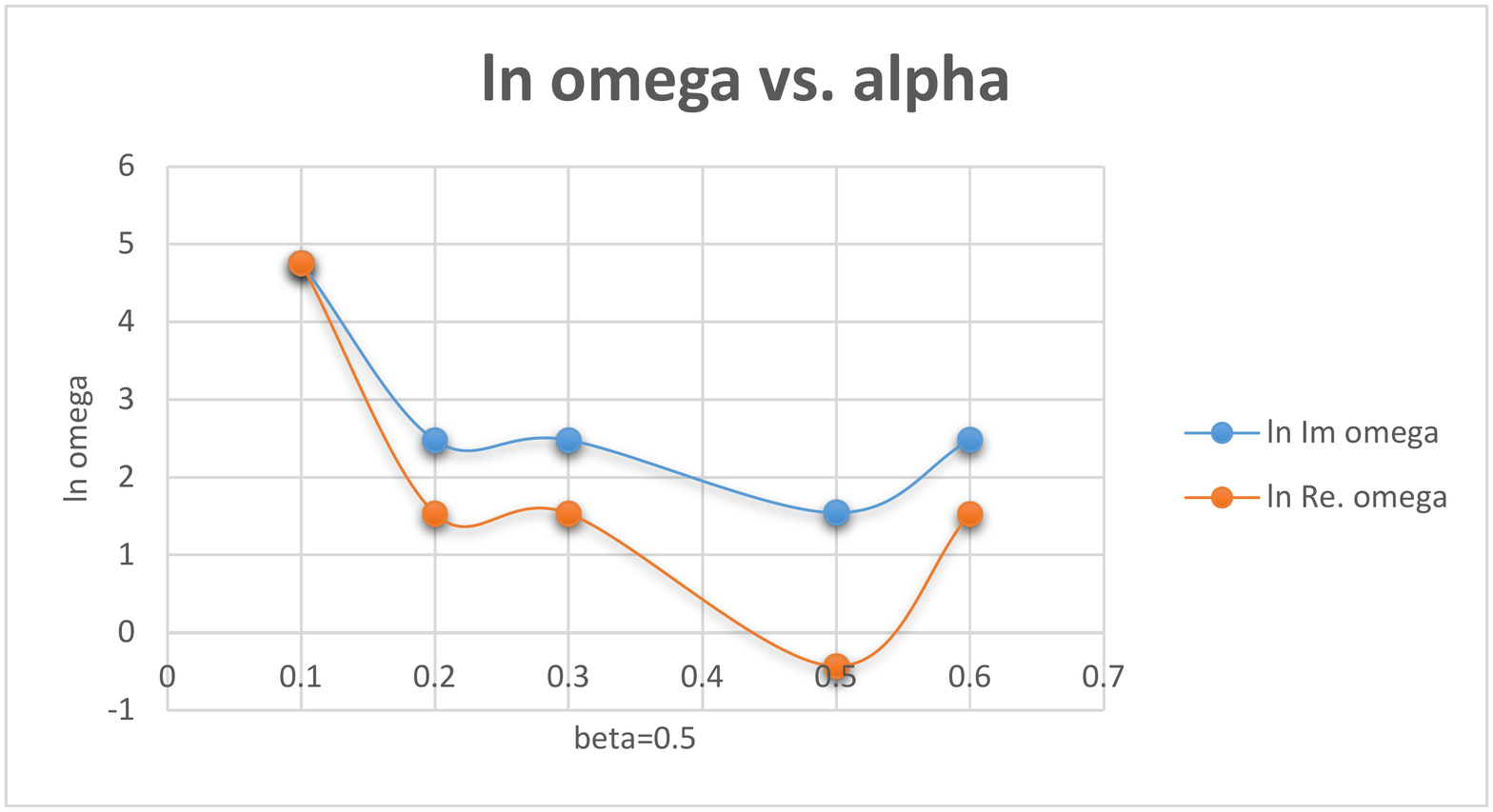}
	\hfill
	\includegraphics[width=.35\textwidth,origin=a,trim=-90 70 200 60, angle=0]{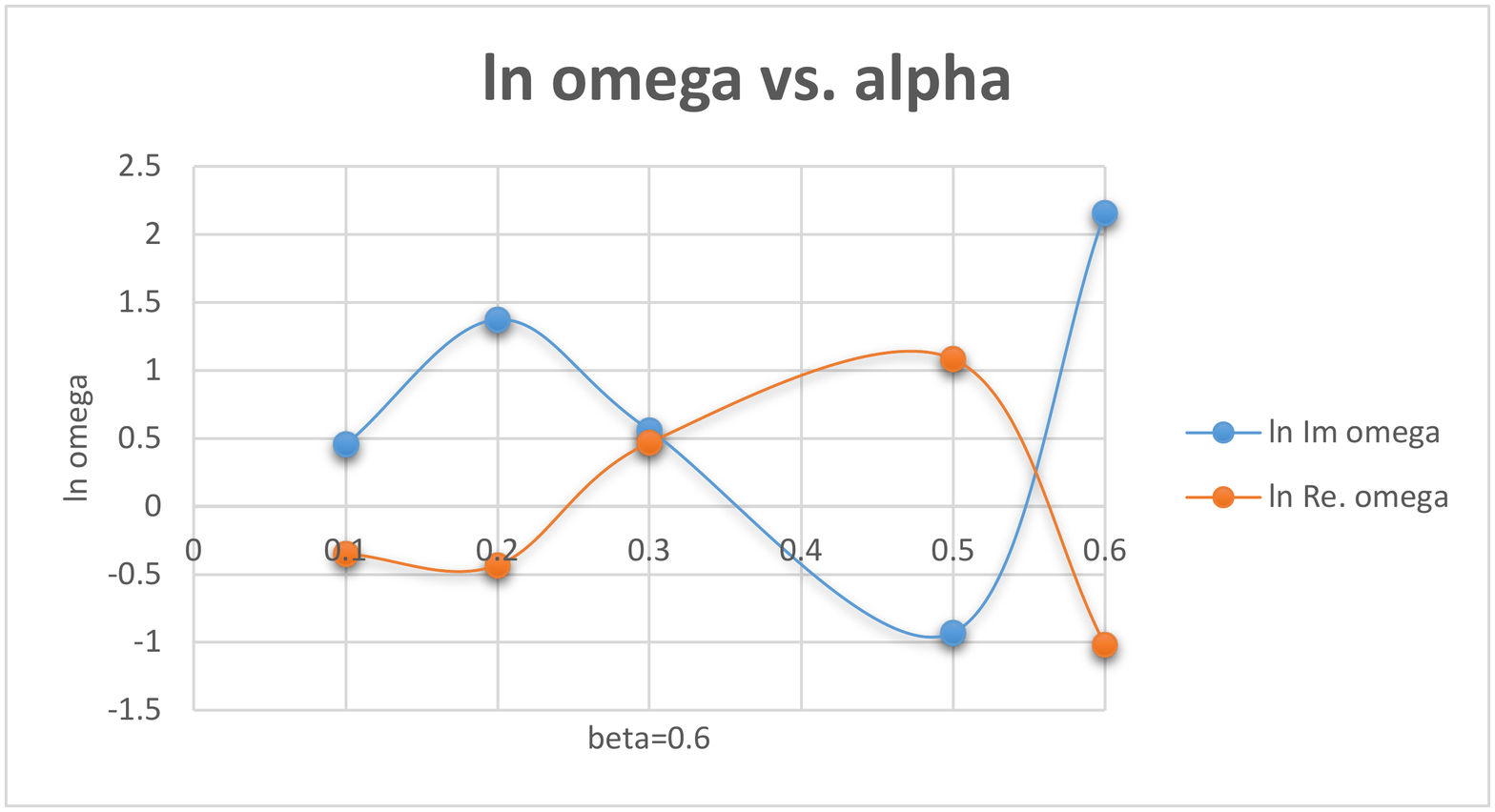}
	\hfill
	\includegraphics[width=.35\textwidth,origin=a,trim=-90 70 200 60, angle=0]{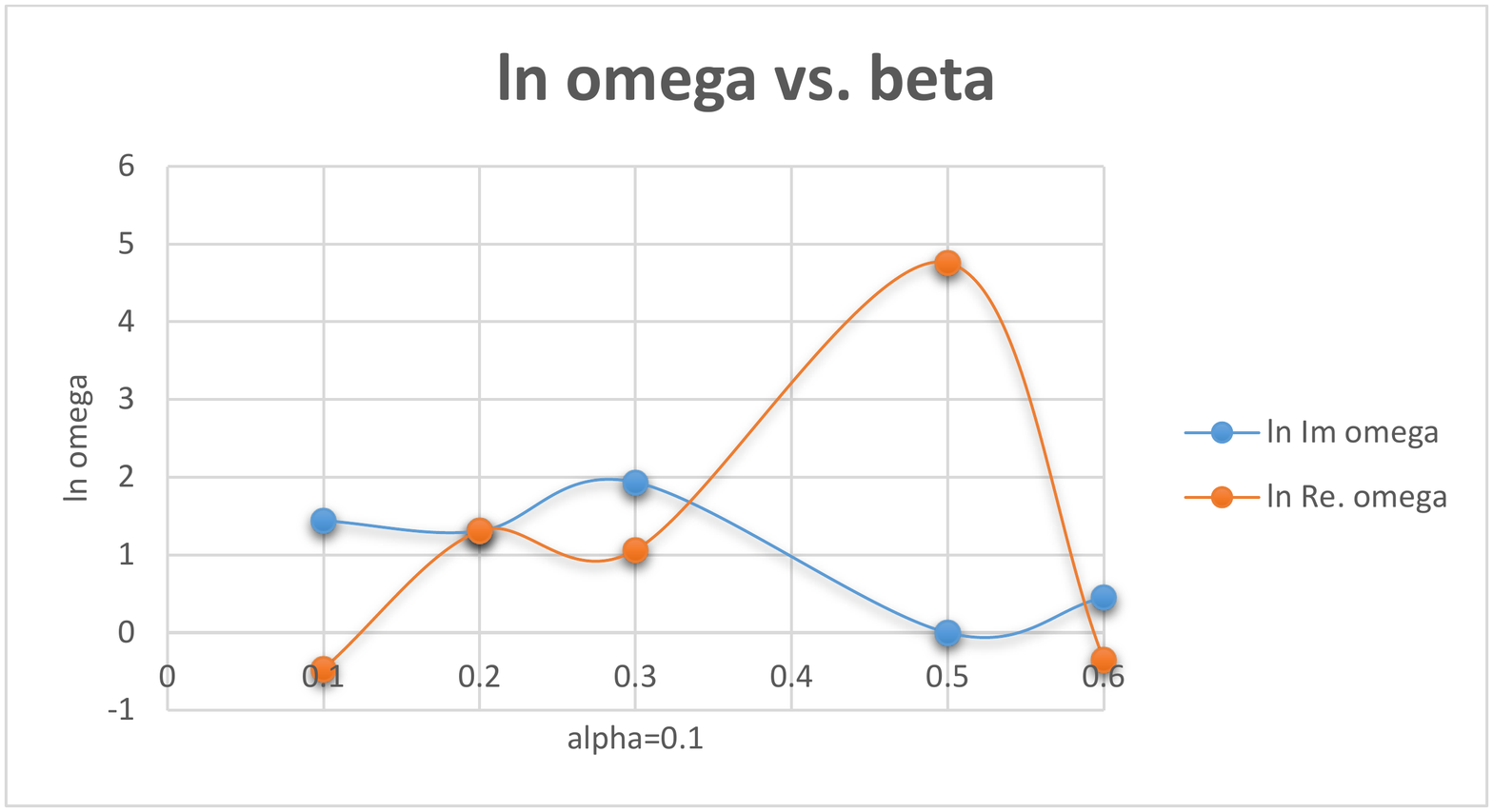}
	\hfill
	\includegraphics[width=.35\textwidth,origin=a,trim=-90 70 220 60, angle=0]{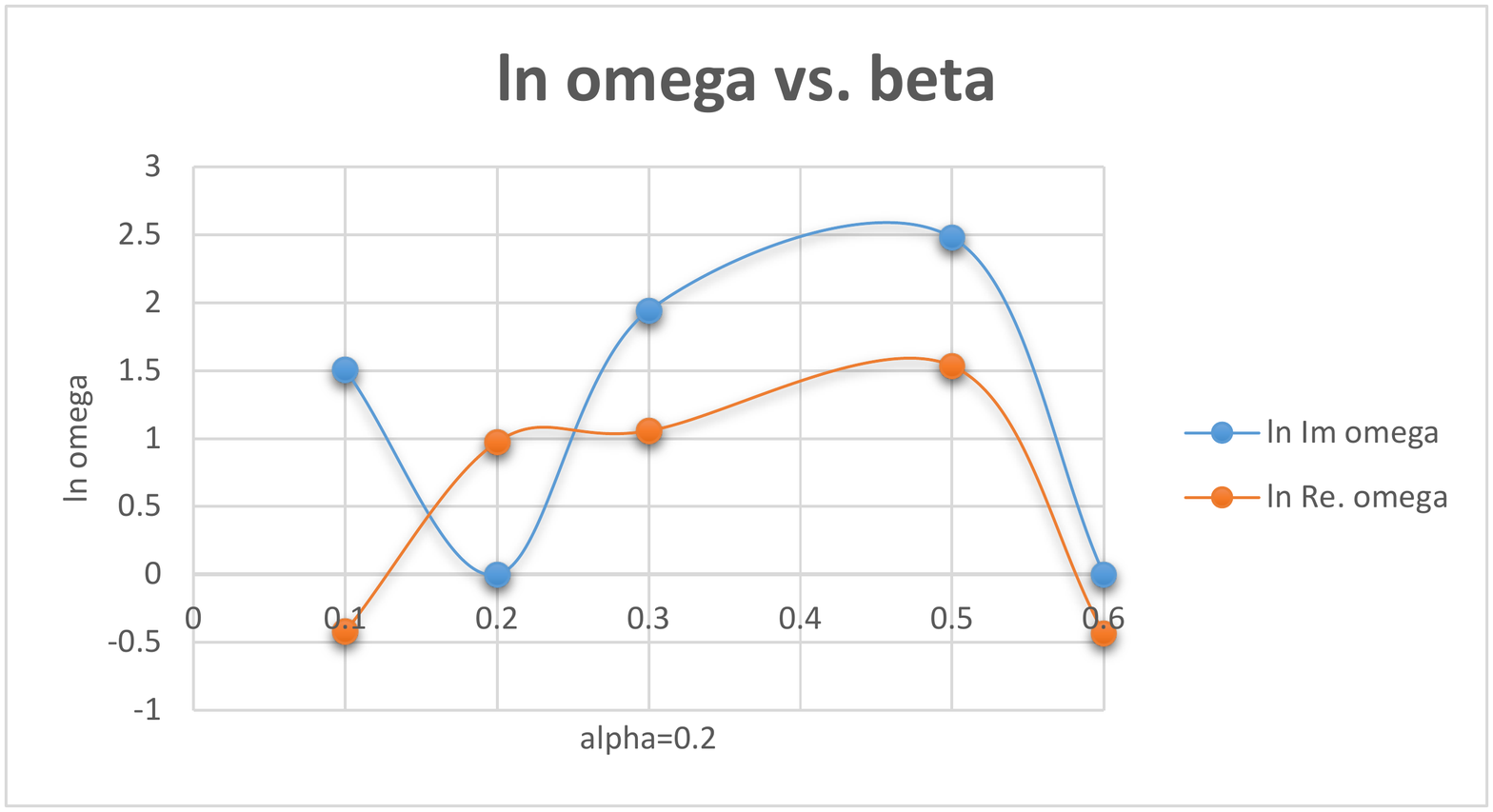}
	\hfill
	\includegraphics[width=.35\textwidth,origin=a,trim=-90 70 220 60, angle=0]{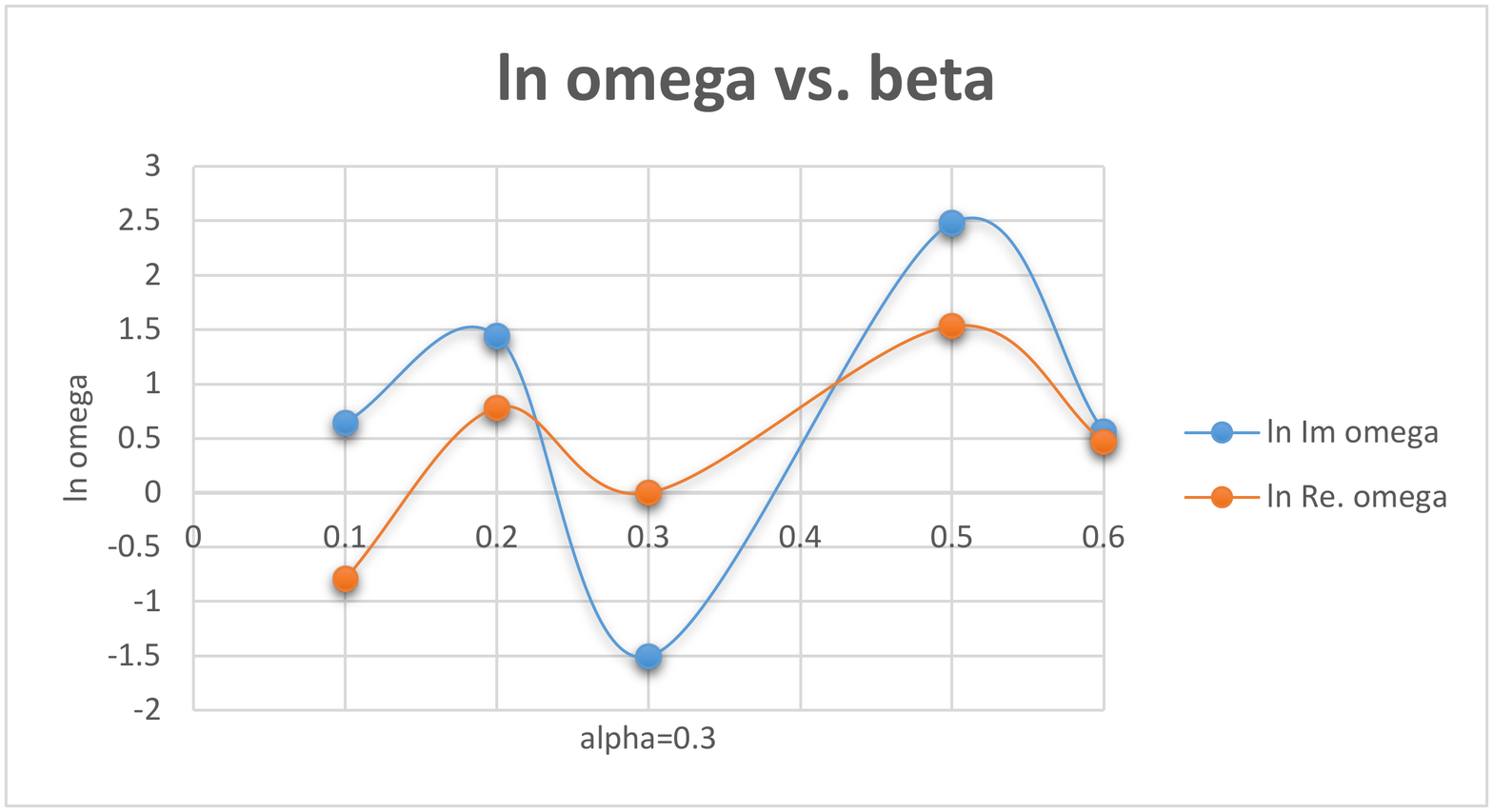}
	\hfill
	\includegraphics[width=.35\textwidth,origin=a,trim=-90 70 200 60, angle=0]{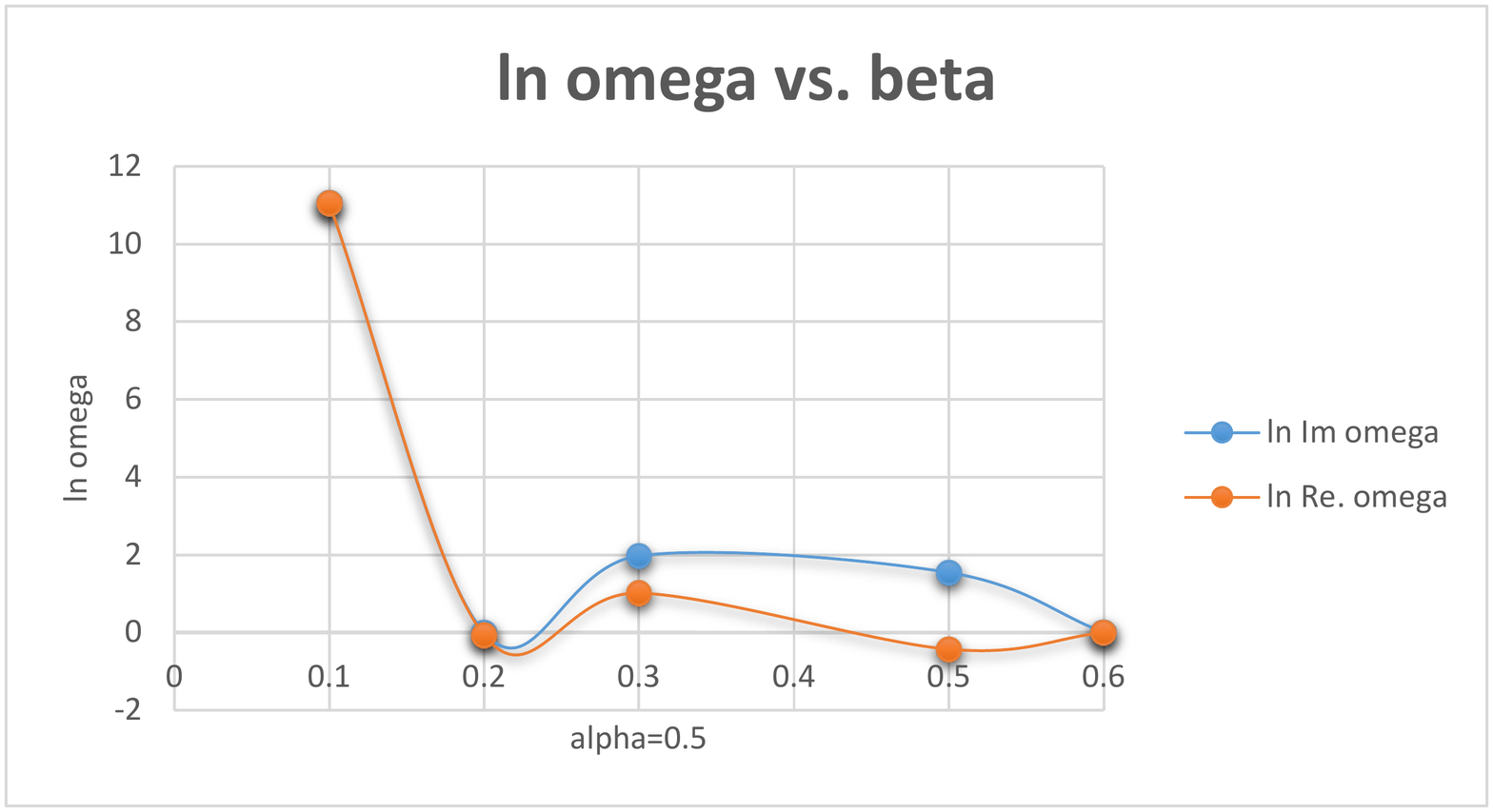}
	\hfill
	\includegraphics[width=.35\textwidth,origin=a,trim=-90 70 200 60, angle=0]{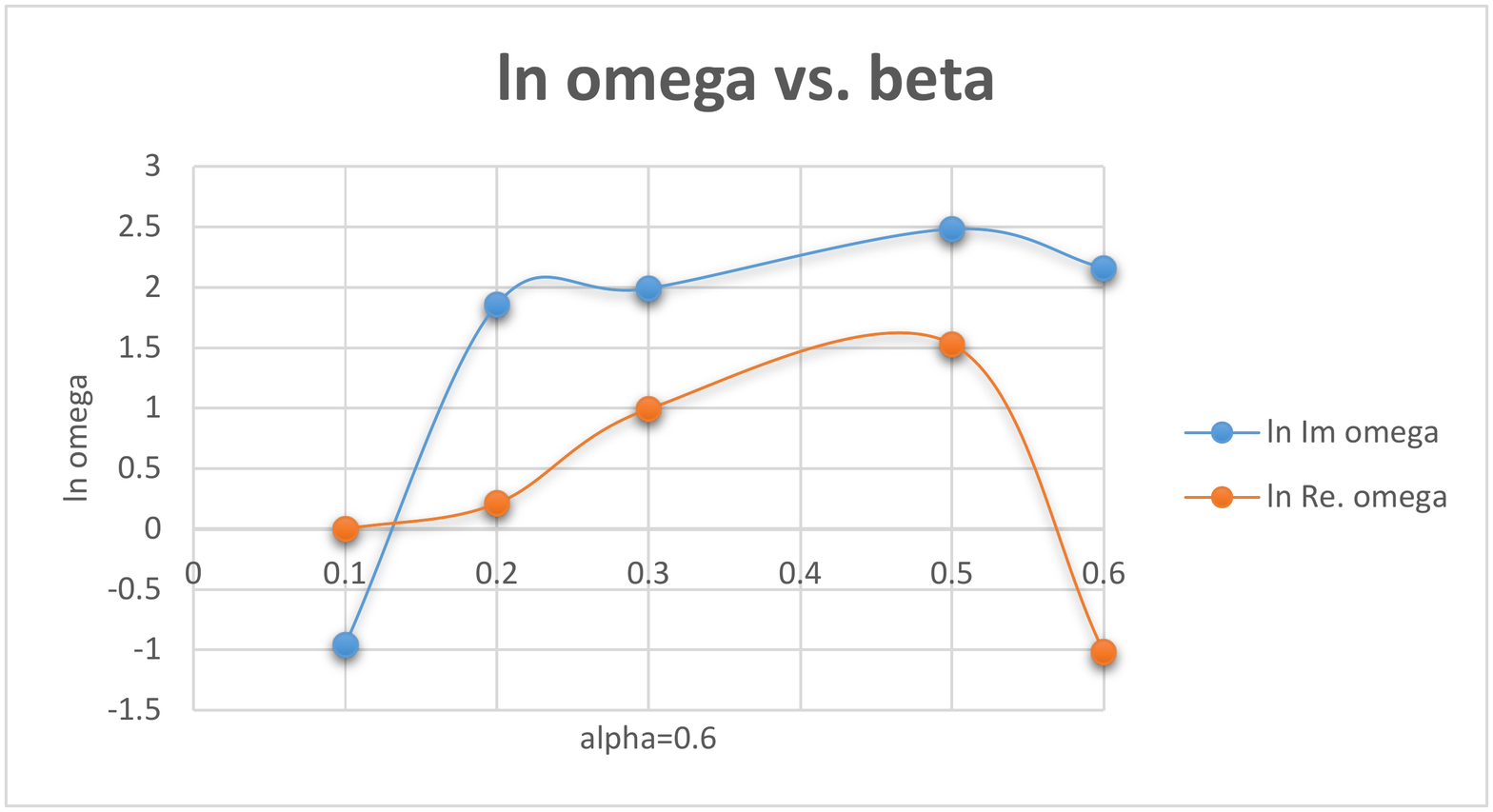}
	\hfill
	
	\caption{\label{fig:i} The Logarithm of real part and imaginary part of the QNM of asymptotically dS black holes as function of $\alpha$ and $\beta$.}
\end{figure}
\begin{figure}[tbp] \label{5}
	\centering 
	\includegraphics[width=.5\textwidth,origin=a,trim=-10 70 100 100,]{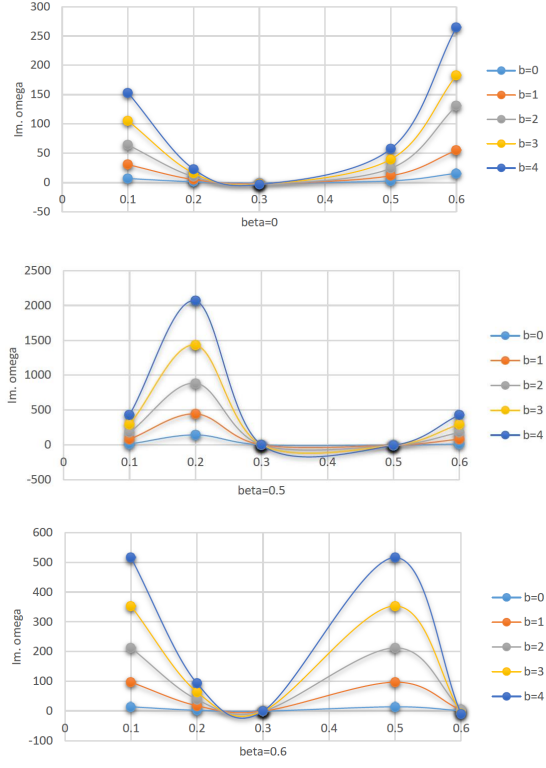}
	\caption{\label{fig:i} The imaginary part of the $\omega$ vs. $\alpha$ for asymptotically AdS black hole in Lovelock background for $n = 6$, $\beta = 0, 0.5, 0.6$, $L = 0$, and $\mu = 1$.}
\end{figure}

\begin{figure}[tbp] \label{6}
	\centering 
	\includegraphics[width=.5\textwidth,origin=a,trim=-10 90 100 100,]{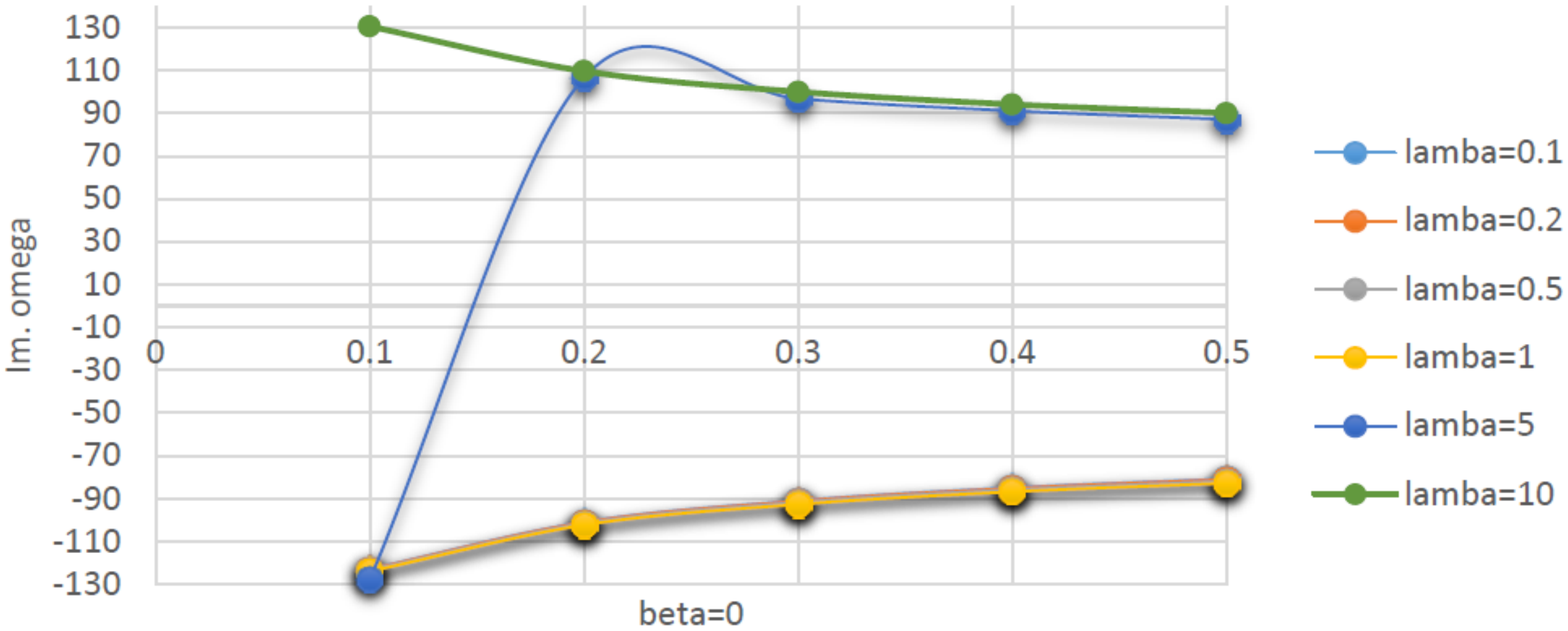}
	\caption{\label{fig:i} The quasinormal mode of the $\omega$ vs. $\alpha$ for asymptotically dS black hole in Lovelock background for $n = 6$, $\beta = 0$, $L = 2$, $\mu = 50$.}
\end{figure}
\begin{figure}[tbp] \label{7}
	\centering 
	\includegraphics[width=.5\textwidth,origin=a,trim=-10 10 100 100,]{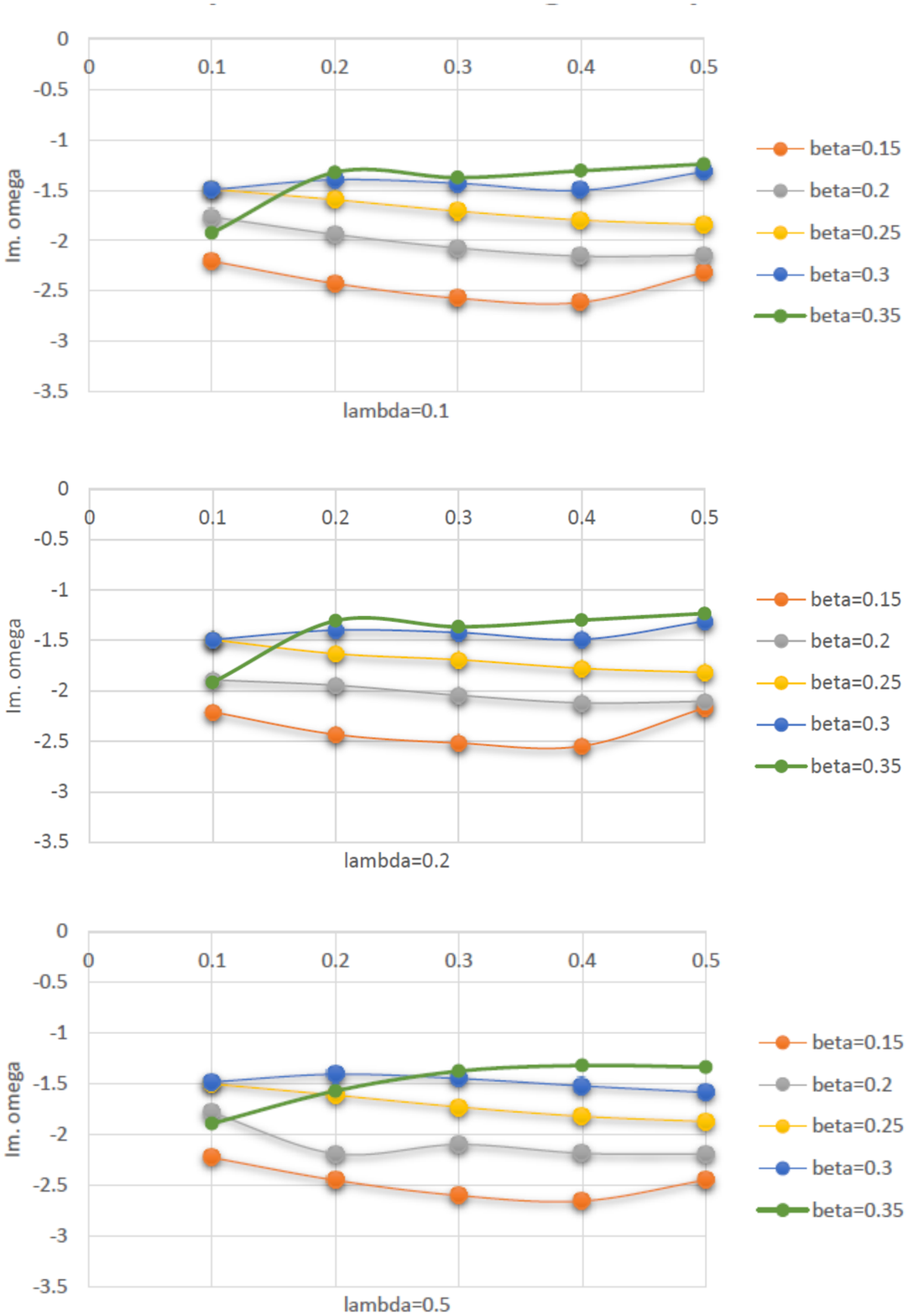}
	\caption{\label{fig:i} The quasinormal mode of the $\omega$ vs. $\alpha$ for asymptotically dS black hole in Lovelock background for $n = 6$, $\Lambda = 0.1,0.2,0.5$, $L = 2$, $\mu = 50$ .}
\end{figure}
\begin{figure}[tbp] \label{8}
	\centering 
	\includegraphics[width=.5\textwidth,origin=a,trim=-10 10 100 100,]{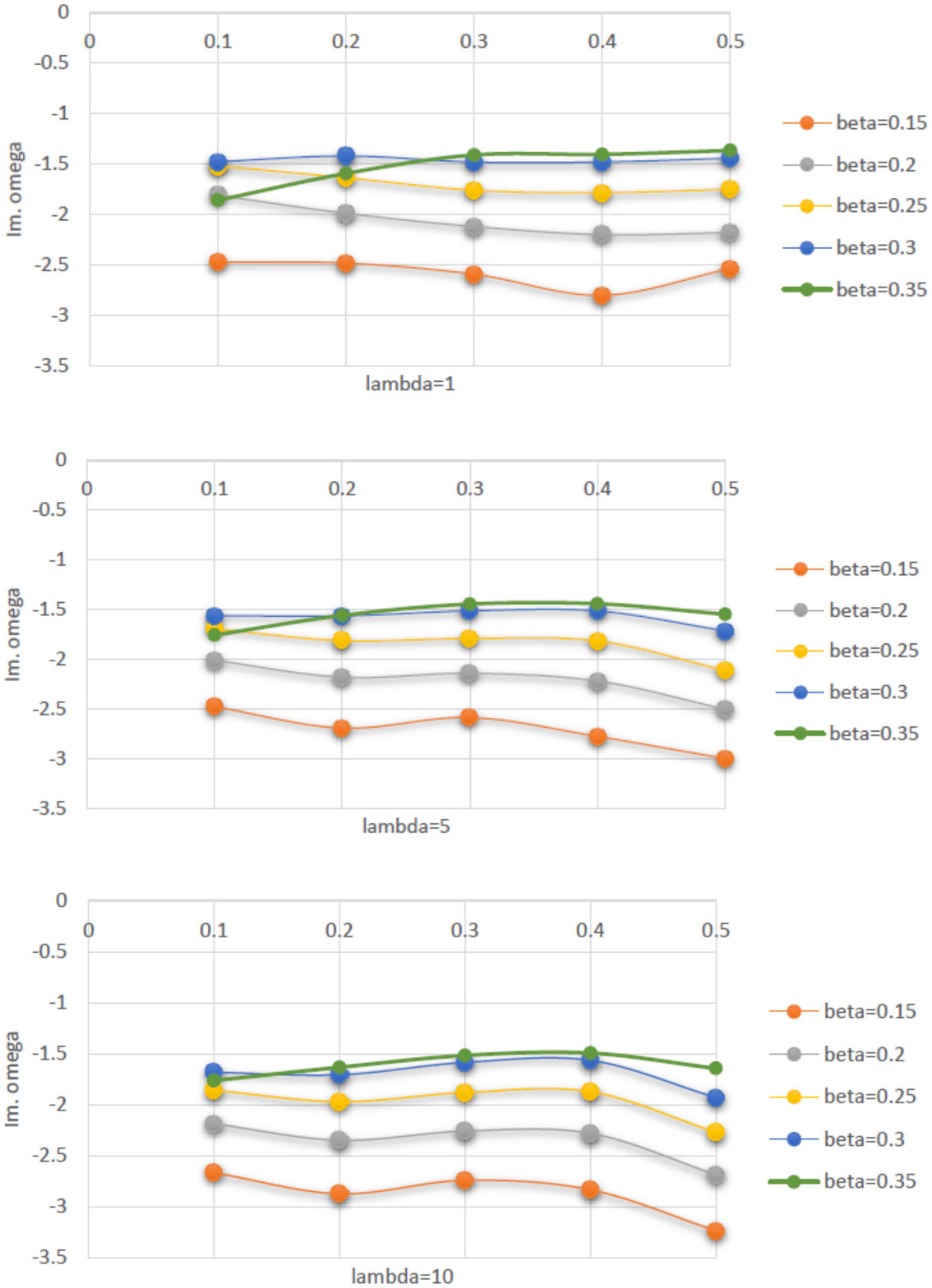}
	\caption{\label{fig:i} The quasinormal mode of the $\omega$ vs. $\alpha$ for asymptotically dS black hole in Lovelock background for $n = 6$, $\Lambda = 1,5,10$, $L = 2$, $\mu = 50$ .}
\end{figure}
\begin{figure}[tbp] \label{9}
	\centering 
	\includegraphics[width=.5\textwidth,origin=a,trim=-10 70 100 100,]{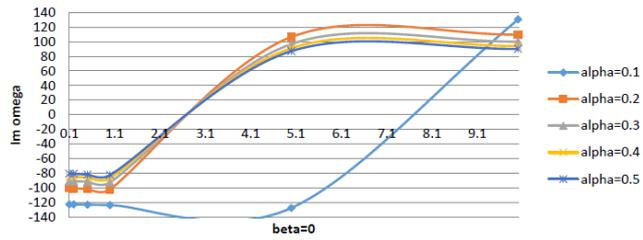}
	\caption{\label{fig:i} The imaginary part of the $\omega$ vs. $\Lambda$ for asymptotically dS black hole in Lovelock background for $n = 6$, $\beta = 0$, $L = 2$, and $\mu = 50$.}
\end{figure} 
\begin{figure}[tbp] \label{10}
	\centering 
	\includegraphics[width=.5\textwidth,origin=a,trim=-10 10 100 100,]{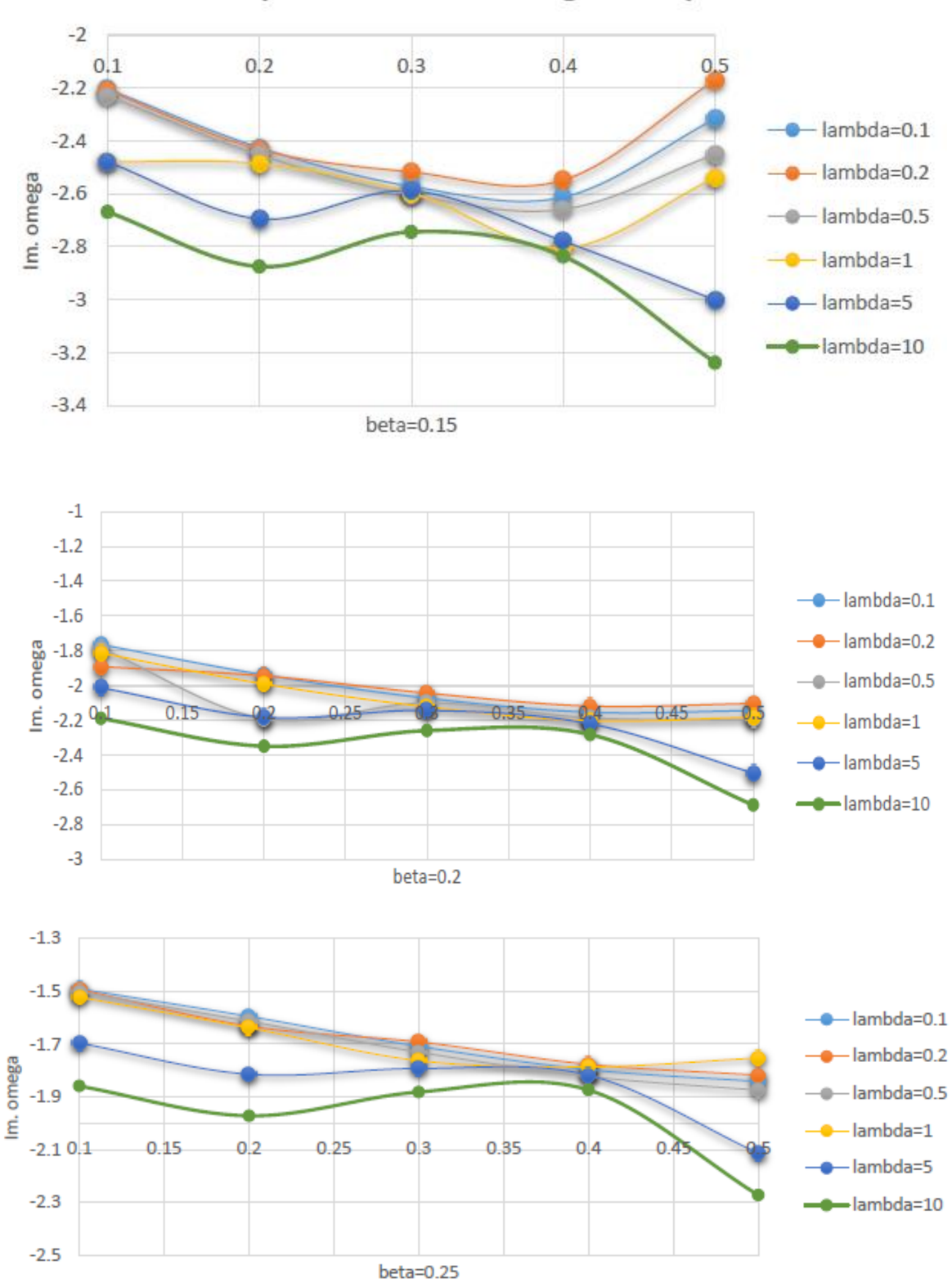}
	\caption{\label{fig:i} The quasinormal mode of the $\omega$ vs. $\alpha$ for asymptotically dS black hole in Lovelock background for $n = 6$, $\beta = 0.15,0.2,0.25$, $L = 2$, $\mu = 50$ .}
\end{figure}
\begin{figure}[tbp] \label{11}
	\centering 
	\includegraphics[width=.5\textwidth,origin=a,trim=-10 180 100 100,]{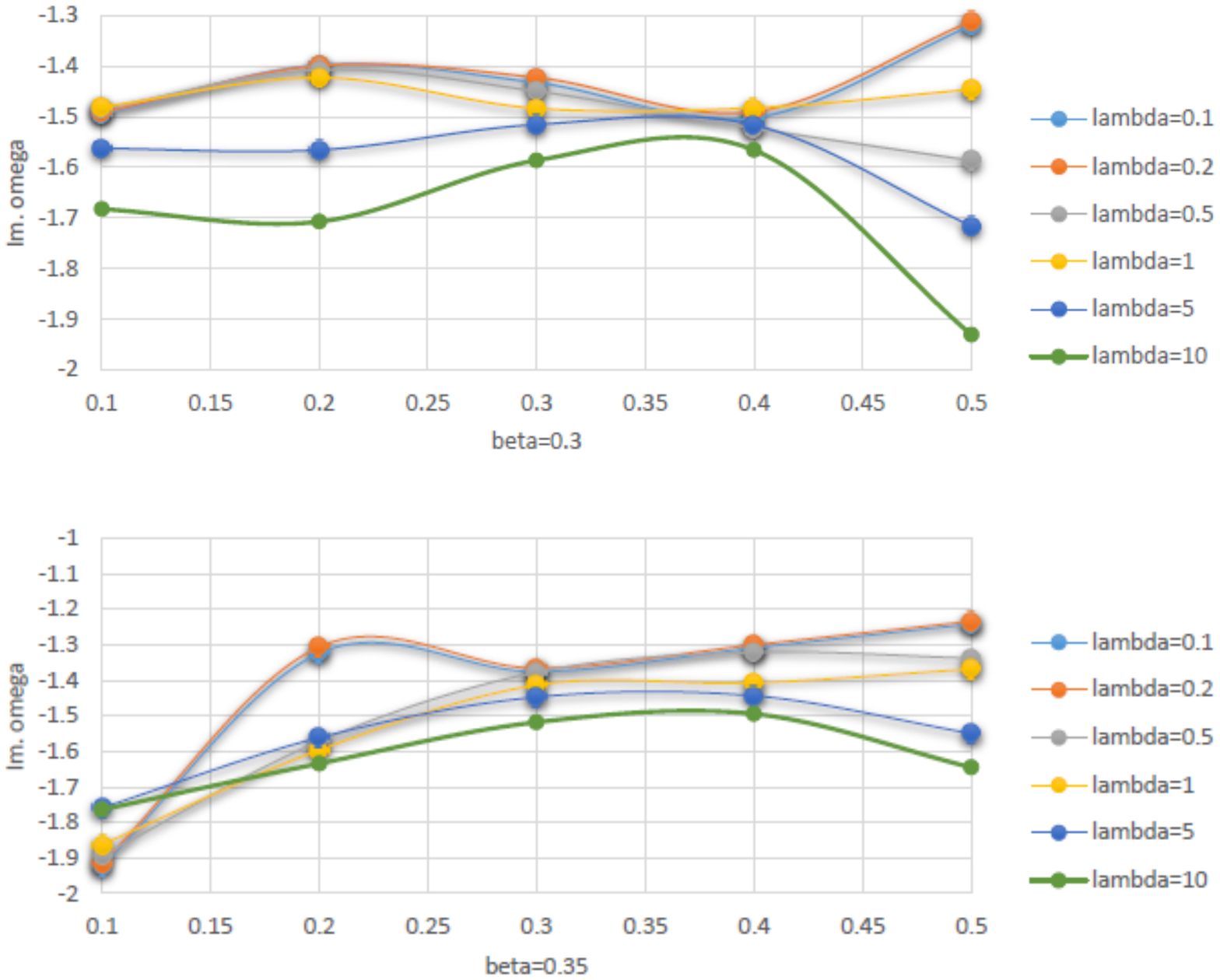}
	\caption{\label{fig:i} The quasinormal mode of the $\omega$ vs. $\alpha$ for asymptotically dS black hole in Lovelock background for $n = 6$, $\beta = 0.3,0.35$, $L = 2$, $\mu = 50$ .}
\end{figure}
\begin{figure}[tbp] \label{12}
	\centering 
	\includegraphics[width=.5\textwidth,origin=a,trim=-10 10 100 100,]{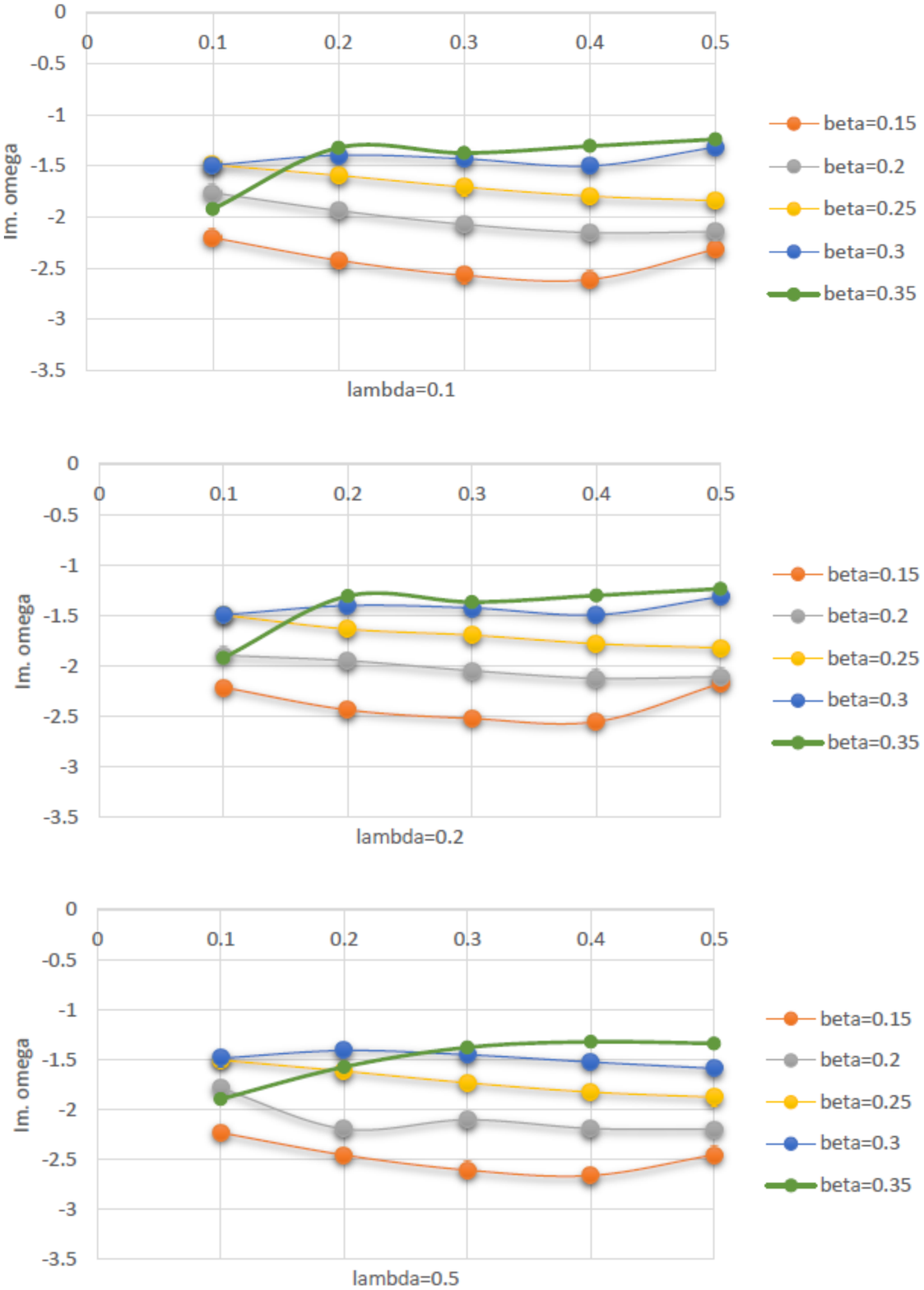}
	\caption{\label{fig:i} The quasinormal mode of the $\omega$ vs. $\alpha$ for asymptotically dS black hole in Lovelock background for $n = 6$, $\Lambda = 0.1,0.2, 0.5$, $L = 2$, $\mu = 50$.}
\end{figure}
\begin{figure}[tbp] \label{13}
	\centering 
	\includegraphics[width=.5\textwidth,origin=a,trim=-10 10 100 100,]{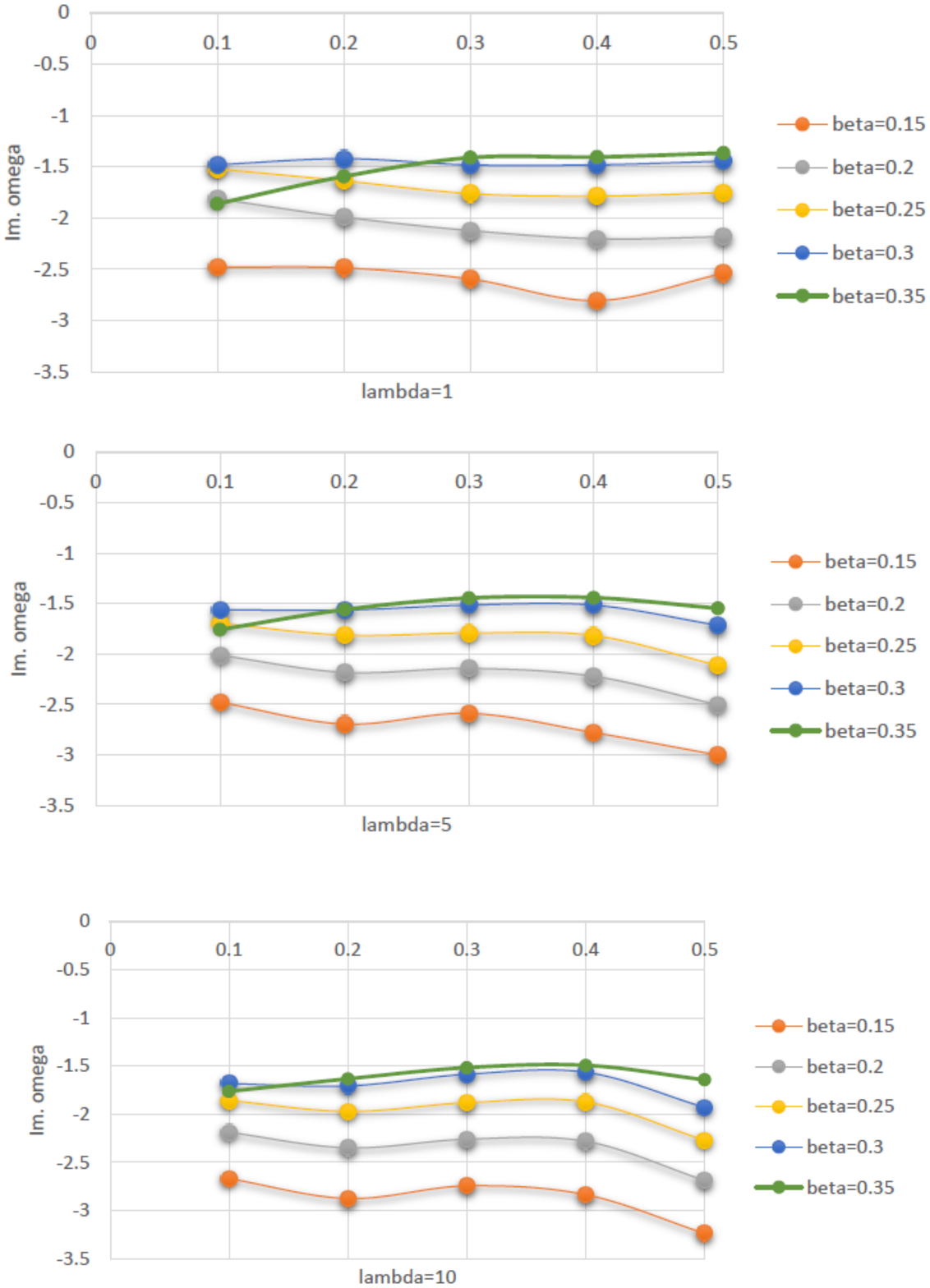}
	\caption{\label{fig:i} The quasinormal mode of the $\omega$ vs. $\alpha$ for asymptotically dS black hole in Lovelock background for $n = 6$, $\Lambda = 1,5,10$, $L = 2$, $\mu = 50$.}
\end{figure}

The tortoise coordinate ${r^*}$ is defined on the interval $\left( { - \infty ,\infty } \right)$ in such a way that the event horizon corresponds to ${r^*} =  - \infty$, while $r=\infty$ corresponds to ${r^*} = \infty$. 
Here, the effective potential has the multipeaks or the negative region for some parameters. As far as one needs to consider the effective potential which has the form of potential barrier and is defined positively, we excluded these parameters and the cases with ghost instability. 
Quasinormal modes are obtained by solving equation (3.4) under appropriate boundary condition.To consider boundary condition for an incoming wave at the event horizon, one can find that the effective potential $V\left( r \right) \to 0$ for $r \to {r_ + }$. The effective potential vanishes, $V\left( r \right) = 0$, in case of an outgoing wave at infinity, as the black hole tends to the Minkowski spacetime at infinity. Therefore, one can write the boundary condition for incoming and outgoing waves as 
\begin{equation}
\left\{ \begin{array}{l}
\psi \left( r \right) \sim {e^{ - i\omega {r^*}}}\begin{array}{*{20}{c}}
{\begin{array}{*{20}{c}}
	{}&{}&{}
	\end{array}}&{as\begin{array}{*{20}{c}}
	{}&{}&{r \to {r_ + }\begin{array}{*{20}{c}}
		{}&{}&{\left( {{r^*} \to  - \infty } \right)}
		\end{array}}
	\end{array}}
\end{array}\\
\psi \left( r \right) \sim {e^{ + i\omega {r^*}}}\begin{array}{*{20}{c}}
{\begin{array}{*{20}{c}}
	{}&{}&{}
	\end{array}}&{as\begin{array}{*{20}{c}}
	{}&{}&{r \to \infty \begin{array}{*{20}{c}}
		{}&{}&{\left( {{r^*} \to  + \infty } \right)}
		\end{array}}
	\end{array}}
\end{array}
\end{array} \right.
\end{equation} 
\\To find the quasinormal frequencies for the massless scalar field in Lovelock black hole spacetime whose effective potential has the form of a potential barrier like that of formula (3.4), one can use the sixth order WKB approximation which was first derived by Schutz and Will \cite{schutz}. This method has extended by Iyer and Will \cite{iyer} to the third  order and then extended to sixth WKB order by Konoplya \cite{konoplya}. 
In this case, when ${\left[ {V\left( r \right) - {\omega ^2}} \right]_{\max }} \ll \left[ {{\omega ^2} - V\left( { \pm \infty } \right)} \right]$, WKB has very good accuracy, i.e. if solutions of equation ${\omega ^2} - V\left( r \right) = 0$ are very close to each other as two turning points and the total energy, ${\omega ^2} - V\left( r \right)$ is expanded to the Taylor series near the maximum of the effective potential near by the turning points.
The sixth order WKB method of the quasinormal mode calculation is governed by the following
\begin{equation} \label{14}
\frac{{i\left( {{\omega ^2} - V\left( {{r_0}} \right)} \right)}}{{\sqrt { - 2V''\left( {{r_0}} \right)} }} + \sum\limits_{i = 2}^6 {{\gamma _i} = b + \frac{1}{2}}
\end{equation} 
where $V\left( {{r_0}} \right)$ is the height and $V''\left( {{r_0}} \right)$ is the second derivative with respect to the tortoise coordinate of the potential at the maximum. Here $b$ is the overtone number. In this case, ${\gamma _i}$ can be found as constants coming from the second up to sixth order WKB correction \cite{konoplya,iyer}.
In order to generate the related effective potential, one can obtain the effective potential for $n = 6$ using the equations (2.7), (2.8), and (3.2) as follows
\begin{equation} \label{14}
\begin{array}{l}
V\left( r \right) = \left( {1 - \frac{{\sqrt[3]{\varsigma }}}{{420{\kern 1pt} \beta {\kern 1pt} r}} + \frac{{7{\kern 1pt} {r^5}\left( {5{\kern 1pt} {\alpha ^2} - 6{\kern 1pt} \beta } \right)}}{{12{\kern 1pt} \beta {\kern 1pt} \sqrt[3]{\varsigma }}} - 1/12{\kern 1pt} \frac{{\alpha {\kern 1pt} {r^2}}}{\beta }} \right) \times \\
\left( \begin{array}{l}
6{\kern 1pt} \frac{1}{{{r^2}}}\left( {1 - \frac{{\sqrt[3]{{\varsigma {\kern 1pt} {r^2}}}}}{{420{\kern 1pt} \beta {\kern 1pt} r}} + \frac{{7{\kern 1pt} {r^5}\left( {5{\kern 1pt} {\alpha ^2} - 6{\kern 1pt} \beta } \right)}}{{12{\kern 1pt} \beta {\kern 1pt} \sqrt[3]{\varsigma }}} - 1/12{\kern 1pt} \frac{{\alpha {\kern 1pt} {r^2}}}{\beta }} \right)\\
+ 3{\kern 1pt} \frac{1}{r}\left( \begin{array}{l}
- \frac{{{r^2}}}{{1260}}\left( {617400{\kern 1pt} \Lambda {\kern 1pt} {\beta ^2}{r^6} - 300125{\kern 1pt} {\alpha ^3}{r^6} + 540225{\kern 1pt} {r^6}\alpha {\kern 1pt} \beta  + \frac{{735{\kern 1pt} \sqrt 5 \beta {\kern 1pt} \chi }}{{2{\kern 1pt} \zeta }}} \right)\\
+ 2{\kern 1pt} {\left( \begin{array}{l}
	88200{\kern 1pt} \Lambda {\beta ^2}{r^7}{\varsigma ^9} - 42875{\kern 1pt} {\alpha ^3}{r^7}{\varsigma ^{10}} + 77175{\kern 1pt} \alpha \beta {r^7}{\varsigma ^9}\\
	+ 735{\kern 1pt} \sqrt 5 \zeta \beta {\varsigma ^2} + 926100{\kern 1pt} {\beta ^2}{\varsigma ^3}\mu \varsigma 
	\end{array} \right)^{ - 2/3}}{\beta ^{ - 1}}\\
+ \frac{{\sqrt[3]{\varsigma }}}{{420{\kern 1pt} \beta {\kern 1pt} {r^2}}} + \frac{{35{\kern 1pt} {r^4}\left( {5{\kern 1pt} {\alpha ^2} - 6{\kern 1pt} \beta } \right)}}{{12{\kern 1pt} \beta {\kern 1pt} \sqrt[3]{\varsigma }}} - \frac{{7{\kern 1pt} {r^5}\left( {5{\kern 1pt} {\alpha ^2} - 6{\kern 1pt} \beta } \right)}}{{36{\kern 1pt} \beta {\kern 1pt} {\varsigma ^{4/3}}}} \times \\
\left( \begin{array}{l}
\left( {617400{\kern 1pt} \Lambda {\kern 1pt} {\beta ^2}{r^6} - 300125{\kern 1pt} {\alpha ^3}{r^6} + 540225{\kern 1pt} {r^6}\alpha {\kern 1pt} \beta  + \frac{{735{\kern 1pt} \sqrt 5 \beta \chi }}{{2{\kern 1pt} \zeta }}} \right){r^2}\\
+ 2{\kern 1pt} \left( {88200{\kern 1pt} \Lambda {\kern 1pt} {\beta ^2}{r^7} - 42875{\kern 1pt} {\alpha ^3}{r^7} + 77175{\kern 1pt} {r^7}\alpha {\kern 1pt} \beta  + 735{\kern 1pt} \sqrt 5 \zeta {\kern 1pt} \beta  + 926100{\kern 1pt} {\beta ^2}\mu } \right)r
\end{array} \right)\\
- 1/6{\kern 1pt} \frac{{\alpha {\kern 1pt} r}}{\beta }
\end{array} \right)\\
+ \frac{{L\left( {L + 5} \right)}}{{{r^2}}}
\end{array} \right)
\end{array}
\end{equation} 
where
\begin{equation} \label{15}
\begin{array}{l}
\zeta  = {\left( \begin{array}{l}
	2880{\kern 1pt} {\Lambda ^2}{\beta ^2}{r^{14}} - 2800{\kern 1pt} \Lambda {\kern 1pt} {\alpha ^3}{r^{14}} + 5040{\kern 1pt} \Lambda {\kern 1pt} \alpha {\kern 1pt} \beta {\kern 1pt} {r^{14}} - 735{\kern 1pt} {\alpha ^2}{r^{14}} + 1176{\kern 1pt} \beta {\kern 1pt} {r^{14}}\\
	+ 60480{\kern 1pt} \Lambda {\kern 1pt} {\beta ^2}\mu {\kern 1pt} {r^7} - 29400{\kern 1pt} {\alpha ^3}\mu {\kern 1pt} {r^7} + 52920{\kern 1pt} \alpha {\kern 1pt} \beta {\kern 1pt} \mu {\kern 1pt} {r^7} + 317520{\kern 1pt} {\beta ^2}{\mu ^2}
	\end{array} \right)^{{1 \mathord{\left/
				{\vphantom {1 2}} \right.
				\kern-\nulldelimiterspace} 2}}}\\
\varsigma  = \left( {88200{\kern 1pt} \Lambda {\kern 1pt} {\beta ^2}{r^7} - 42875{\kern 1pt} {\alpha ^3}{r^7} + 77175{\kern 1pt} {r^7}\alpha {\kern 1pt} \beta  + 735{\kern 1pt} \sqrt 5 \zeta {\kern 1pt} \beta  + 926100{\kern 1pt} {\beta ^2}\mu } \right){r^2}\\
\chi  = 40320{\kern 1pt} {\Lambda ^2}{\beta ^2}{r^{13}} - 39200{\kern 1pt} \Lambda {\kern 1pt} {\alpha ^3}{r^{13}} + 70560{\kern 1pt} \Lambda {\kern 1pt} \alpha {\kern 1pt} \beta {\kern 1pt} {r^{13}} - 10290{\kern 1pt} {\alpha ^2}{r^{13}} + 16464{\kern 1pt} \beta {\kern 1pt} {r^{13}}\\
+ 423360{\kern 1pt} \Lambda {\kern 1pt} {\beta ^2}\mu {\kern 1pt} {r^6} - 205800{\kern 1pt} {\alpha ^3}\mu {\kern 1pt} {r^6} + 370440{\kern 1pt} \alpha {\kern 1pt} \beta {\kern 1pt} \mu {\kern 1pt} {r^6}
\end{array}
\end{equation}
Which depends on the value of $r$, integration constant related to $ADM$ mass, $\mu$,the second and third orders of Lovelock gravity coefficients, $\alpha$ and $\beta$, cosmological constant and angular momentum as well. It is straightforward to obtain the effective potential for other dimensions of the spacetime such as $n = 3,4,5$. We would like to note that, in case of $n = 4$, the third order of Lovelock gravity is no longer valid to the master equation (2.8) and it reduces to Einstein-Gauss-Bonnet black holes. 
\\Figure 1 represent the typical variation of the effective potential $V\left( r\right)$ with respect to $r$ in presence of the Lovelock correction coefficients, $\alpha$ and $\beta$. The different plots are drawn for different values of $\alpha$ and $\beta$. It shows the potential variation for different $\beta$ as the third order coupling constant of Lovelock theory.
One can deduce that the $\beta$ parameter increases the height of the potential barrier governed by the effective potential. Figure 1 shows the cases $\alpha=0.1$, $\alpha=0.3$ with $\beta=0.5$ as two example of bad potential which has been taken out from the calculation as mentioned before. However, Figure 2 shows a well behaved potential which is in the form of potential barrier.  
\\We now apply the WKB method extended to the sixth order adopted to Lovelock theory \cite{soda} to compute the frequencies of the QNMs of the metric perturbations that obey equation (3.4). Within the WKB approach in order to take into account of the second and third order of the Lovelock gravitational theory, we use $\alpha \ge 0$ and $\beta \ge 0$. Here, first we set $\mu = 1$, $L=0$ and $\Lambda=\pm1$. Then, we computed the QNM frequencies for different overtone from $b=0$ to $b=4$.
In Figure 3 the imaginary parts of the QNM frequencies are presented for the asymptotically dS black holes in Lovelock background as a function of $\alpha$ and $\beta$ for $n = 6$ dimension and different values of the overtone including the fundamental state. We assume the perturbations depend on ${e^{ - i\omega t}}$ according to the equation (3.2). Therefore, to have damping, the imaginary part of $\omega$ must be negative which could be satisfied by setting $\alpha = \beta$ (See Figure 3 and Table I).
The calculation is done for different positive values of the second and third order of Lovelock coefficients (see Table I). Figure 3 shows the behavior of the imaginary part of the QNM frequencies for the asymptotically dS Lovelock black hole as a function of the Lovelock coupling constants for $8$ dimensional spacetime. 
In this case, the different values of overtones including the fundamental state is presented and increasing the overtone causes more variations of the imaginary part of the QN frequencies for all values of the second and third orders of Lovelock coupling constants.
Figure 3 shows for $\beta = 0$, the real part of the quasinormal mode frequency increases with $\alpha$. But the imaginary part shows a different kind of behavior, the damping decreases as $\alpha$ is increased from some lower values to some minimum and then the damping increases as we go to larger values of $\alpha$. Obviously, taking into account of $\beta > 0$ changes the behavior of the real part of the QNMs of Lovelock black holes. Figure 3 also shows no difference between the imaginary part and real part of the QNMs for $\beta = 0$ and $\alpha > 0.3$. It is shown that variation of the imaginary and real parts of the QNMs of the massless scalar field in asymptotically dS background in Lovelock theory are related to the second and third order of the coupling constants of Lovelock, $\alpha$ and $\beta$. Table 1 shows different values of the quasinormal modes of asymptotically dS Lovelock black hole for $L = 0$. In the same manner, one can obtain the effective potential for asymptotically AdS black hole in Lovelock background and compute the quasinormal mode using WKB approximation. In this case, the imaginary part of the QNMs in fundamental state of overtone has small variation as a function of $\alpha$ and $\beta$ (see Figures 5). Here, increasing the overtone causes more variations of the imaginary part of the QN frequencies for all values of the second and third orders of Lovelock coupling constants which confirm that the value of overtone not to be more than angular momentum. Therefore, in next step, we compute the quasinormal mode of Lovelock black holes for fundamental overtone with $L = 2$.  
\\In this manner, the quasinormal mode of the $\omega$ with respect to the $\alpha$ for asymptotically dS black hole in Lovelock background are well-defined (see Figure 6, 7, and 8). In this case we calculate the quasinormal modes of black hole with $\alpha = 0.1$ to $0.5$ with interval $0.1$ for different values of $\beta$ from $0.15$ to $0.35$ with interval $0.05$ in case of $\Lambda = 0.1$. 
\\In this case, we use the same method to compute the quasinormal modes for different values of cosmological constant such as $0.1$, $0.2$, $0.5$, $1$, $5$, and $10$ for asymptotically dS black holes and $-0.1$, $-0.2$, $-0.5$, $-1$, $-5$, and $-10$ for asymptotically AdS black holes in Lovelock gravity. In case of $\beta = 0$ which is Gauss-Bonnet black hole case, the quasinormal modes are well-behaved except $\Lambda = 0.1$ and $\alpha = 0.1$ (Figure 9). In case, increasing $\beta$ does not cause variation on imaginary part of the $\omega$ for different values of $\alpha$ except in case of $\Lambda = 10$ which has more variation compare to the other values of $\Lambda$ (see Figures 10, 11, 12, and 13).  
\\In this way, the imaginary part of the $\omega$ is not well-behaved for $\alpha = 0.1$ in case of asymptotically AdS black hole with $\beta = 0$. However, there is no result for $\Lambda > -5$ and in case of $\alpha > 0.4$ for $\Lambda = 1$. In this case, the quasinormal modes of the AdS black holes in Lovelock background diverge with increasing cosmological constants (see Table III). Here one can refer to the Table (I)II as well for computed quasinormal modes of the (A)dS black holes in Lovelock background.     
\section{Conclusion}
In this paper, we have investigated quasinormal modes of massless scalar field of the asymptotically dS black holes in Lovelock background up to eight dimensions. We found that the effective potential $V$ depends only on the value of $r$, integration constant related to $ADM$ mass, $\mu$, and Lovelock second and third order coefficients, $\alpha$ and $\beta$, respectively. We have computed the quasinormal frequencies spectrum of the asymptotically (A)dS black holes in Lovelock gravity theory, using the sixth order of WKB approximation. It is shown that the quasinormal spectrum depends on the second and third order of Lovelock coupling constant parameters, $\alpha$ and $\beta$. The parameters $\alpha$ and $\beta$ are given values from 0.1 to 0.6. The calculation is done for different values of $\alpha$ and $\beta$ for $L = 0$ (see Tables I) and $L = 2$ (Tables II and III). We compare the real part and imaginary part of the quasinormal frequencies and their variations respect to second and third order of the Lovelock coupling constants (Figure 4). We also have computed the quasinormal modes for different overtone including the fundamental state. We observed that increasing the overtone causes more variations of the imaginary part of the quasinormal frequencies for all values of the second and third order of Lovelock coupling constants, $\alpha$ and $\beta$ in case of $L = 0$. Figures 5 shows the variation of imaginary part of the quasinormal modes of the $\omega$ with respect of the parameters $\alpha$ and $\beta$ for $L = 0$. We found in case, if we set $L$ smaller than overtone the quasinormal mode spectrum would diverge. Therefore, the calculation was done for the case $L = 2$ for fundamental overtone, $b = 0$ as well. One can find the behavior of the spectrum in case of the $L = 2$ in Figure 6 specifically for Gasuss-Bonnet dS black holes. Figures 7, 8, and 10 show the quasinormal mode of the $\omega$ with respect to the $\alpha$ for asymptotically dS black hole in Lovelock Back ground. We found remarkable features of the quasinormal frequencies of Lovelock (A)dS black holes which
depend on the coefficients of the Lovelock terms, the species of perturbations, and dimensions of the spacetime. Indeed, increasing the coefficients of the third order Lovelock term causes increasing of the real part of quasinormal frequencies which is in the same manner of asymptotically flat black holes in Lovelock back ground computed in Ref. \cite{soda}. In this study, we found, decreasing the cosmological constant in AdS spactime causes divergence in quasinormal spectrum. Therefore, although, Lovelock theory has a rich structure of AdS vacua, the 6th order of the WKB approximation is not precise to compute the poles of the retarded correlators of the thermal Super Yang Mills theory using AdS/CFT correspondence by investigating quasin normal modes of asymptotically AdS spherical static black holes in Lovelock gravity.  
\begin{acknowledgments}
	N. Abbasvandi would like to thank Prof. Jiro Soda for his insightful discussion and support. The paper is supported by Universiti Kebangsaan Malaysia (Grant No.FRGS/2/2013/ST02/UKM/02/2 for N. Abbasvandi). Shahidan Radiman would like to to acknowledge receiving a grant DPP-2015-036 from UKM.
\end{acknowledgments}


\end{document}